\documentclass[sigconf]{acmart}
\usepackage{amsmath}
\usepackage{graphicx}
\usepackage{multirow}
\usepackage{epstopdf}
\usepackage{wrapfig}
\usepackage{algorithm, algorithmic}
\usepackage{tikz}
\usepackage{bm}
\usepackage{dsfont} 
\usepackage[mathscr]{euscript} 
 \let\mathscr\relax
\usepackage[scr]{rsfso} 
\usepackage{enumitem}
\usetikzlibrary{bayesnet}
\usepackage{enumitem}
\usepackage{caption}
\usepackage{xspace}
\usepackage{bbm}
\usepackage{mathtools} 

\usepackage{cuted} 
\pdfoutput=1

\newcommand*{\function}{\mathord{\mathit{f}}}

\newenvironment{myquote}[1]%
  {\list{}{\leftmargin=#1\rightmargin=#1}\item[]}%
  {\endlist}  

\AtBeginDocument{%
\providecommand\BibTeX{{%
\normalfont B\kern-0.5em{\scshape i\kern-0.25em b}\kern-0.8em\TeX}}}

\copyrightyear{2023}
\acmYear{2023}
\setcopyright{rightsretained}
\acmConference[KDD '23]{Proceedings of the 29th ACM SIGKDD Conference on Knowledge Discovery and Data Mining}{August 6--10, 2023}{Long Beach, CA, USA}
\acmBooktitle{Proceedings of the 29th ACM SIGKDD Conference on Knowledge Discovery and Data Mining (KDD '23), August 6--10, 2023, Long Beach, CA, USA}
\acmDOI{10.1145/3580305.3599471}
\acmISBN{979-8-4007-0103-0/23/08}

\makeatletter
\gdef\@copyrightpermission{
  \begin{minipage}{0.3\columnwidth}
   \href{https://creativecommons.org/licenses/by/4.0/}{\includegraphics[width=0.90\textwidth]{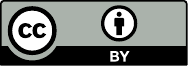}}
  \end{minipage}\hfill
  \begin{minipage}{0.7\columnwidth}
   \href{https://creativecommons.org/licenses/by/4.0/}{This work is licensed under a Creative Commons Attribution International 4.0 License.}
  \end{minipage}
  \vspace{5pt}
}
\makeatother

\begin{document}

\title[Preemptive Detection of Fake Accounts on Social Networks]{Preemptive Detection of Fake Accounts on Social Networks via Multi-Class Preferential Attachment Classifiers}

\author{Adam Breuer}
\email{abreuer@dartmouth.edu}
\affiliation{%
  \institution{Dartmouth}
  \city{Hanover}
  \state{NH}
  \country{USA}
}%

\author{Nazanin Khosravani}
\email{nazaninkt@fb.com}
\affiliation{%
  \institution{Meta}
  \city{Menlo Park}
  \state{CA}
  \country{USA}
}%

\author{Michael Tingley}
\email{tingley@fb.com}
\affiliation{%
  \institution{Meta}
  \city{Menlo Park}
  \state{CA}
  \country{USA}
}%

\author{Bradford Cottel}
\email{bcottel@fb.com}
\affiliation{%
  \institution{Meta}
  \city{Menlo Park}
  \state{CA}
  \country{USA}
}%

\renewcommand{\shortauthors}{Adam Breuer, Nazanin Khosravani, Michael Tingley, \& Bradford Cottel}

\begin{abstract}

In this paper, we describe a new algorithm called \textbf{Pre}ferential \textbf{Attac}hment $\mathbf{k}$-class Classifier (\textsc{PreAttacK}) for detecting fake accounts in a social network. Recently, several algorithms have obtained high accuracy on this problem. However, they have done so by relying on information about fake accounts' friendships or the content they share with others---the very things we seek to prevent. 

\textsc{PreAttacK} represents a significant departure from these approaches. We provide some of the first detailed distributional analyses of how new fake (and real) accounts first attempt to make friends by strategically targeting their initial friend \emph{requests} after joining a major social network (Facebook). We show that even before a new account has made friends or shared content, these initial friend \emph{request} behaviors evoke a natural multi-class extension of the canonical Preferential Attachment model of social network growth. 

We leverage this model to derive a new algorithm, \textsc{PreAttacK}. We prove that in relevant problem instances, \textsc{PreAttacK} near-optimally approximates the posterior probability that a new account is fake under this multi-class Preferential Attachment model of new accounts' (not-yet-answered) friend requests. These are the first provable guarantees for fake account detection that apply to new users, and that do not require strong homophily assumptions.

This principled approach also makes \textsc{PreAttacK} the only algorithm with provable guarantees that obtains state-of-the-art performance at scale on the global Facebook network, allowing it to detect fake accounts before standard methods apply and at lower computational cost.  
Specifically, \textsc{PreAttacK} converges to informative classifications (AUC $\approx$$0.9$) after new accounts send + receive a total of just $20$ not-yet-answered friend requests. 
For comparison, state-of-the-art network-based algorithms do not obtain this performance even after observing additional data on new users' first $100$ friend requests. Thus, unlike mainstream algorithms, PreAttacK \emph{converges before the median new fake account has made a single friendship (i.e. \emph{accepted} friend request)} with a human.

\end{abstract}




\maketitle


\section{Introduction}
\label{intro}
Fake user accounts are the primary source of fake news and other malicious phenomena on social networks such as Facebook and Twitter. Organized campaigns of fake accounts have recently been used to influence public opinion, push propaganda, infiltrate political discourse, manipulate stock markets, steal personal data, and propagate scams ~\citep{bessi2016social, confessore2018follower, chu2010tweeting, shu2017fake, lamb2018felt, thomas2011suspended, stieglitz2017social, shao2017spread, ferrara2016rise, fandos2018facebook, varol2017online, Standards2021}. Detecting these fake accounts and limiting their ability to interact maliciously with humans are core tasks for modern social networks \citep{kozlov2020evaluating, xudeep, breuer2020friend}.

The scale of fake accounts has increased commensurately with the rapid growth of online social networks. In the last year alone, Facebook disabled $6.1$ billion fake accounts---more than double the number of active users on the Facebook network~\citep{Standards2021}. This figure reflects immense recent progress in fake account classification---for example, Facebook disabled the vast majority of these fakes during account registration. Nonetheless, the fraction of \emph{active} social network users who are fake has remained at roughly $4$-$5$\% (for Facebook) or $8$-$15$\% (for Twitter) for the last several years \citep{Standards2021, varol2017online}.

\paragraph{\textbf{The early detection paradox}}
\label{para:paradox}
These active fakes that evade registration-time classifiers and join a social network raise what we call the early detection paradox: \emph{Mainstream algorithms to detect active fake accounts rely on information about their friends or the content they share with others, yet these friendships and shared content are the very things we seek to prevent.} \noindent Our goal in this paper is to design algorithms that overcome this paradox by classifying active fake accounts \emph{before} they make friends or share content.


\begin{figure}
\centering
\includegraphics[height=0.28\textwidth,
bb=0 0 612 540]{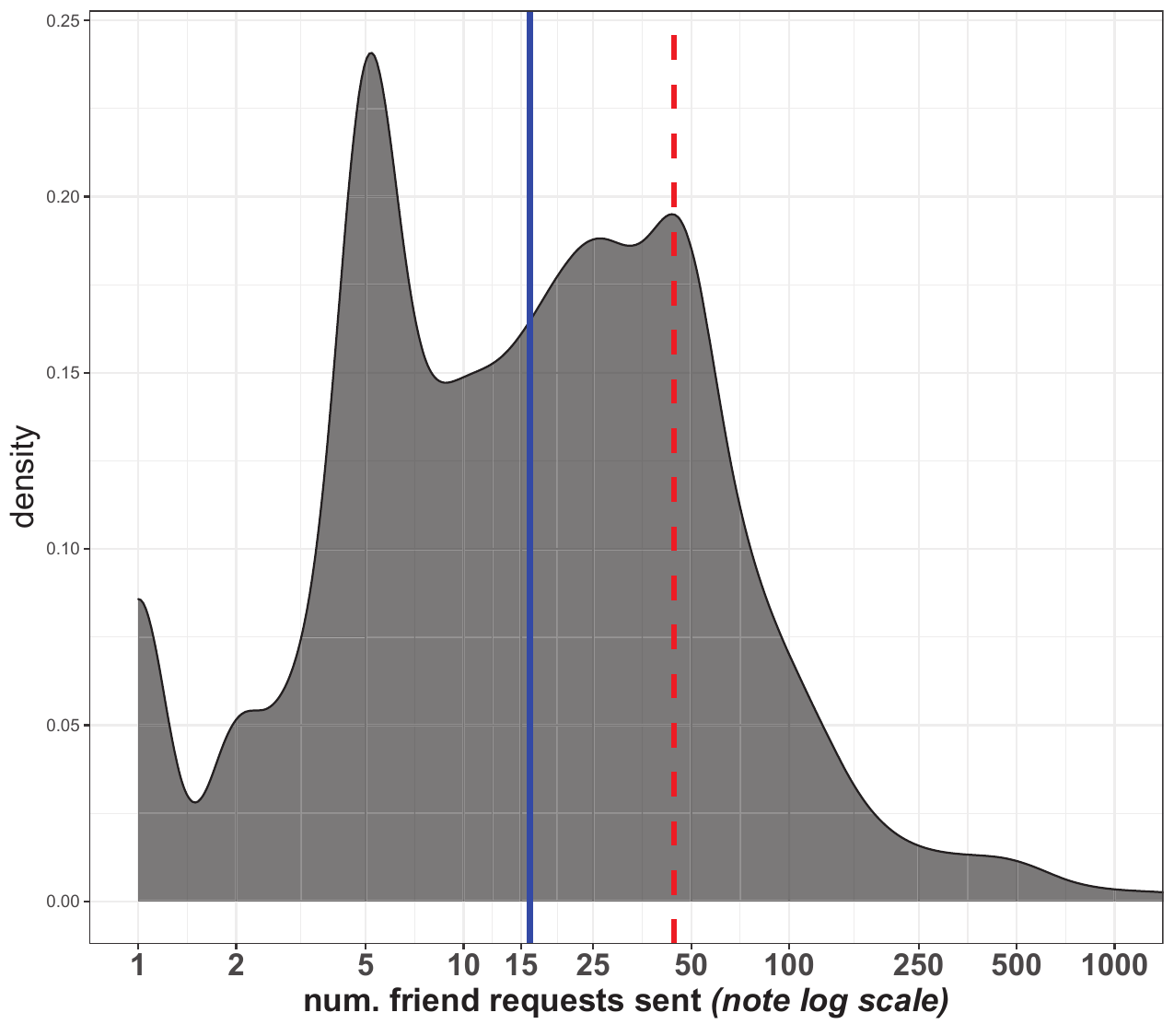}
\caption{Distribs. of counts of Facebook friend requests \emph{sent} by new fake accounts before a single real user accepts any \emph{among new fake accounts who eventually befriend a real user}. Median at blue solid line; mean at dashed red line.}
\Description[Motivating distributions]{} 
\label{fig:motivation_plots1}
\vspace{-1mm}
\end{figure}

\paragraph{\textbf{Recent algorithms}} This paradox is captured by the two mainstream approaches to fake account detection:

\begin{itemize}
    \item \emph{Network-structural algorithms.} Network-structural algorithms classify long-tenured accounts via the \emph{Homophily Assumption}, which states that users \emph{eventually} tend to cluster together (i.e. make the majority of their steady-state friendships) with other users who share their same $\{fake, real\}$ label~\citep{yu2006sybilguard, yang2012sybilrank, gong2014sybilbelief, wang2018structure}. Based on this assumption, network-structural algorithms attempt to propagate a small number of known users' $\{fake, real\}$ labels across the friendship network to unknown users via either Random Walks ~\citep{yu2006sybilguard, yu2008sybillimit, danezis2009sybilinfer, jia2017random, boshmaf2015integro, yang2012sybilrank} or Belief Propagation~\citep{gong2014sybilbelief, gao2018sybilfuse, wang2017sybilscar, wang2018structure, wang2017gang}.\smallskip
    \item \emph{Feature-based classifiers.}
Recently, a variety of research has detected fake accounts in a supervised learning setting. State-of-the-art algorithms such as \textsc{DEC} \citep{xudeep}, \textsc{Jodie} \citep{kumar2019predicting} and \textsc{Ties} \citep{noorshams2020ties} accomplish this via embeddings of tens of thousands of features that capture sophisticated properties of a user's friendship network, such as the average account age of a user's friends-of-friends, or temporal trends in the content a user shares over time~\citep{stein2011facebook, kudugunta2018deep, volkova2017identifying, khaled2018detecting, xudeep, noorshams2020ties, kumar2019predicting}. 
While these algorithms have no theoretic guarantees, they are performant: Facebook now uses them to obtain high quality $\{fake, real\}$ labels (AUC$>$$0.98)$ for virtually all of their long-tenured users~\citep{xudeep, kozlov2020evaluating, breuer2020friend}.
\end{itemize}

Notwithstanding these impressive results, neither approach is ideally suited to the \emph{early detection} of new fake accounts that have not yet made many (or any) friendships: Because such accounts have just passed registration-time feature-based classifiers, they cannot be detected by other feature-based classifiers until their features evolve significantly. Also, many informative features are unknown until after a new user has made several friends or shared content with others. Similarly, it is well-known that mainstream network-structural algorithms do not apply, as their theoretic guarantees rely critically on the Homophily Assumption, which only applies to long-tenured users who have had sufficient `stabilization time' to make the majority of their eventual friendships ~\citep{yang2012sybilrank, boshmaf2015integro, al2017sybilreview, ramalingam2018fakesybilreview, wang2018structure}. For this reason, evaluations of network-structural algorithms have often excluded new users with less than e.g. $1$ to $6$ months of tenure on the social network~\citep{boshmaf2015integro, yang2012sybilrank, boshmaf2015thwarting}. Recent evaluations of these algorithms on the Facebook network suggest they perform poorly (AUC<$0.6$) on new users who have not yet made many friends \citep{breuer2020friend}. 

\begin{figure}
\centering
\includegraphics[height=0.28\textwidth]{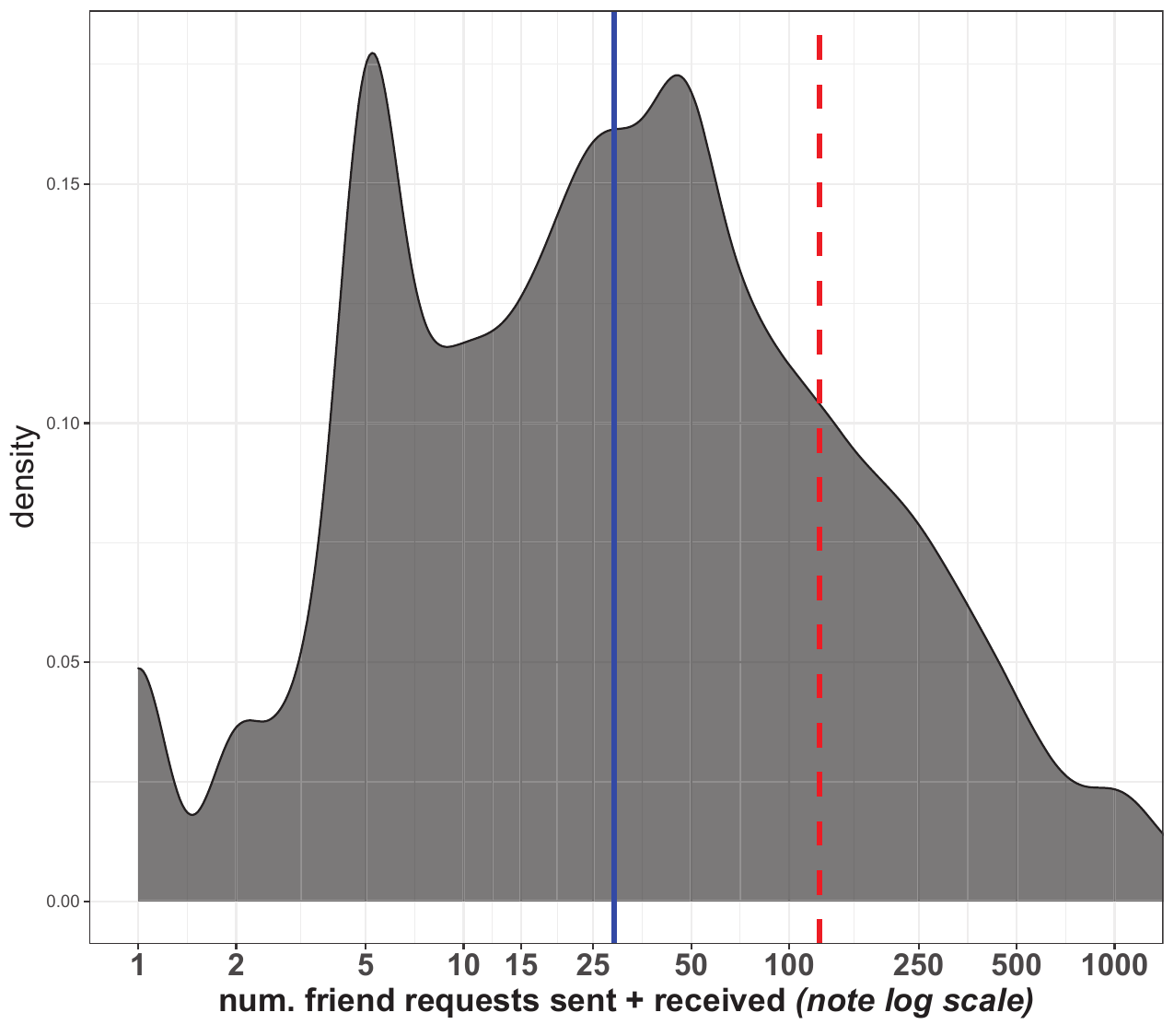}
\caption{Distribs. of counts of friend requests \emph{sent + received}.}
\Description[Motivating distributions]{} 
\label{fig:motivation_plots2}
\vspace{-4mm}
\end{figure}

\paragraph{\textbf{Overcoming the paradox}}  To address this early detection paradox, we use data from the Facebook social network to provide some of the first distributional analyses of how fake (and real) accounts target their friend \emph{requests} after joining a major social network (Figs. 1-4). This focus on friend requests is motivated by the fact that new fake accounts can only meaningfully interact with real users after they have sent friend requests to real users (or received requests from real users) \emph{and} those requests have been seen and accepted.  Fig. \ref{fig:motivation_plots1} shows that \emph{among the subset of new fake accounts that eventually obtain a friendship with a real user,} the median new fake account sends $16$ friend requests before obtaining a single friendship (\emph{accepted} request) with a real user (note log-scale). If we also include the requests these new fake accounts receive from others (Fig. \ref{fig:motivation_plots2}), the count increases to $29$ requests (sent+received).

\smallskip
\begin{myquote}{0.16in}
\emph{Can we leverage this small number of not-yet-answered friend requests to distinguish new fake accounts from new real users?}
\end{myquote}

\vspace{-0.4mm}

\begin{figure}
\centering
\includegraphics[height=0.3\textwidth]{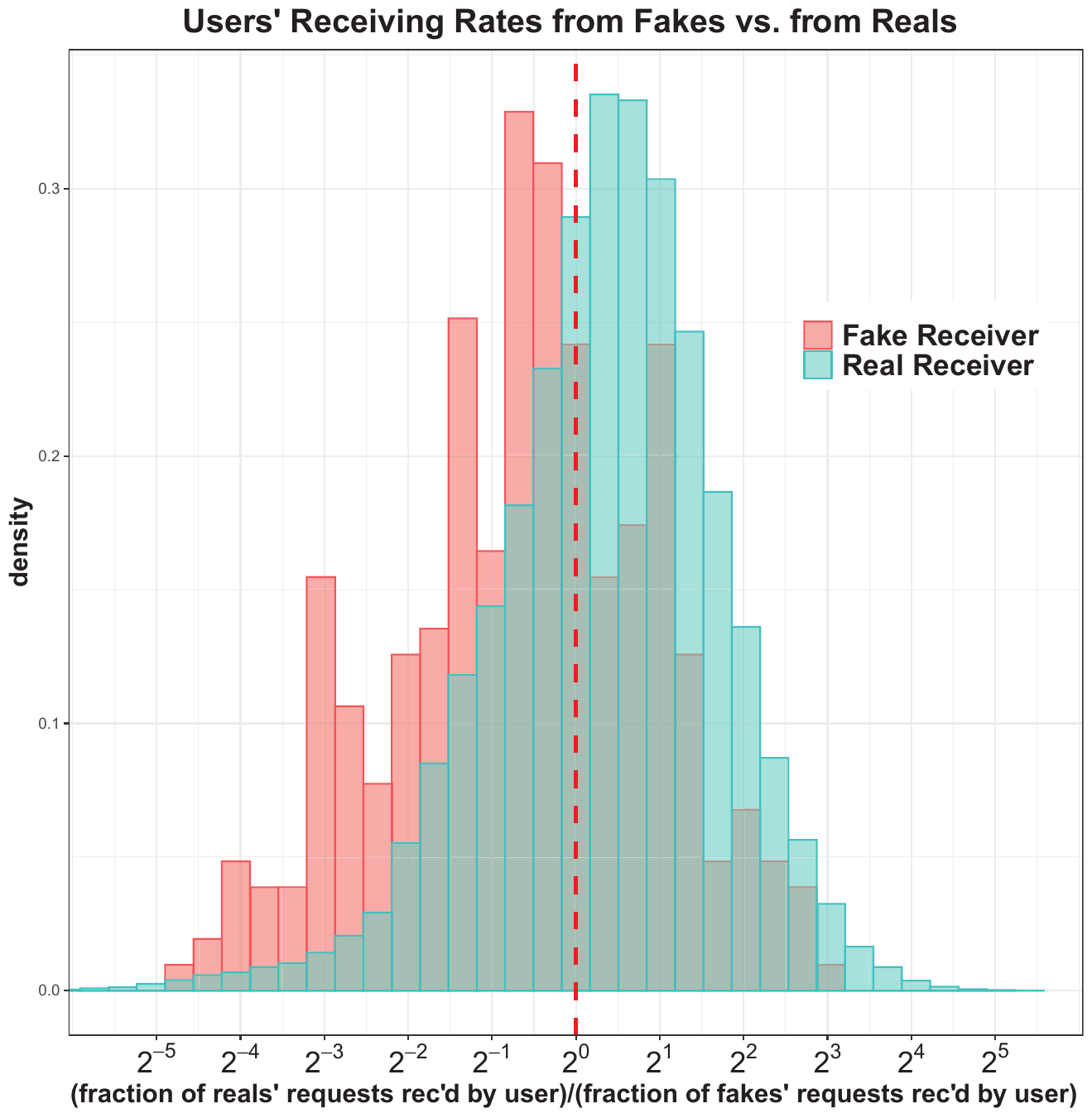}
\caption{Mass right of $x$=$1$ line represents $\{fake, real\}$ Facebook users who \emph{receive} disproportionately more of real users’ requests vs. fakes’ requests by the factor on the $x$ axis.}
\Description[Observed Preferential Attachment Probabilities]{} 
\label{fig:PArateplots1}
\vspace{-5mm}
\end{figure}

\begin{figure}
\centering
\includegraphics[height=0.3\textwidth]{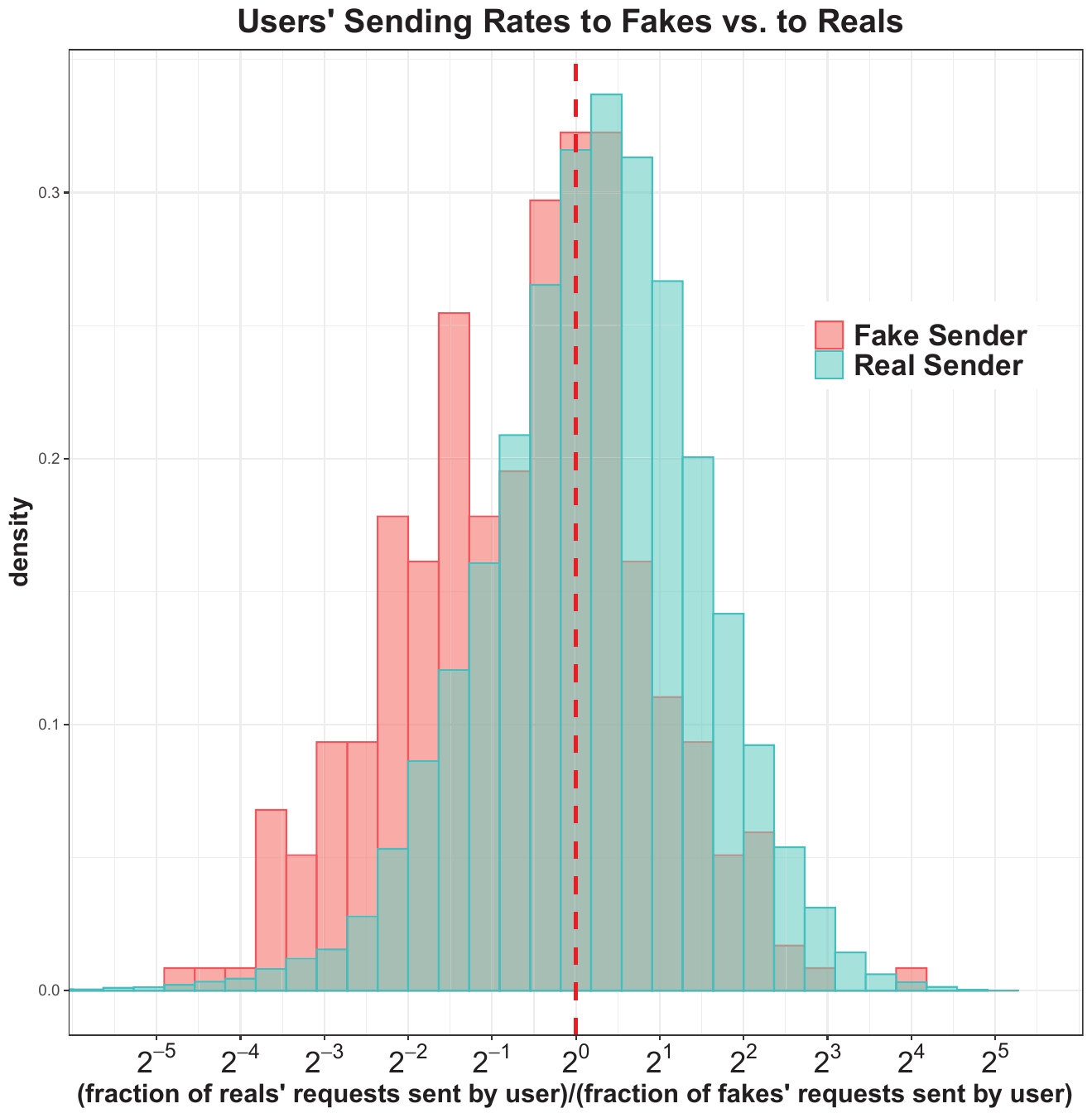}
\caption{ Mass right of $x$=$1$ line represents FB users who \emph{send} disproportionately more requests to real users vs. to fakes.}
\Description[Observed Preferential Attachment Probabilities]{} 
\label{fig:PArateplots2}
\vspace{-6mm}
\end{figure}

\paragraph{\textbf{$k$-Class Directed Preferential Attachment model ($k$CDPA)}} On the Facebook network, we observe that while fake and real users do differ slightly as a class in terms of the degree to which they send and receive requests from fakes and reals (\emph{red vs. blue distributions in Figs. \ref{fig:PArateplots1} and \ref{fig:PArateplots2}, next pg.---see also Sec. \ref{sec:PreAttacKpp}}), these class-level differences are small in comparison to individual-level differences (\emph{spread of distributions}). Specifically, some users are exponentially more likely to request (or be requested by) a real user (\emph{mass right of red lines in Figs. \ref{fig:PArateplots1} and  \ref{fig:PArateplots2}, resp.}); Other users are exponentially more likely to request (or be requested by) a fake account (\emph{mass left of red lines}).

This observation evokes the canonical Preferential Attachment (PA, i.e. rich-get-richer) generative model of social network growth \citep{price1976general, albert2002statistical, bollobas2011degree, barabasi1999emergence, backstrom2012four}. In a traditional PA model, each new user joins a social network and sends friend requests to recipients who are selected with probability proportional to the counts of requests that they have already received. This process results in a power-law distribution of users' in-degrees such that a small number of recipients become vastly more popular than others. PA models and their associated dynamic processes continue to motivate a variety of recent results across several machine learning subfields.

\vspace{7mm}
In our problem setting, fake and real users' `preferential attachment' to \emph{different} individuals inspires a natural multi-class extension of the PA model, which we call $k$CDPA:
\begin{itemize}
    \item Suppose we observe an \emph{arbitrary preexisting} directed network of friend requests between existing users. Then, suppose some new fake and real users join this network.
    \item New fakes and reals each send and receive friend requests to/from existing users who are chosen proportional to how many \emph{of the new user's $fake/real$ class} already did so.
\end{itemize}

The $k$CDPA model provides a principled foundation for a classifier that applies to new accounts. Specifically, recent research has highlighted various similar multi-class PA models as a theoretical mechanism for the emergence of homophily in social networks \citep{, pmlr-v38-lee15b, avin2015homophily, zhang2018understanding, avin2020mixed, nettasinghe2021directed}. As such, our $k$CDPA model forms a natural antecedent to standard homophily-based fake account detection methods that are used to detect long-tenured fake accounts.

We emphasize that we use $k$CDPA to model the friend request networks of a small batch of new users; we do \emph{not} assume that the entire network emerged from this process (which would be a far stronger assumption), nor do we assume that the (distinct) network of \emph{accepted} friend requests (i.e. friendships) adheres to PA.

\paragraph{\textbf{Main contribution.}}
Our main result is an algorithm, \textsc{PreAttacK}, that determines the posterior probability that a new user is a fake account based on the $k$CDPA model of her (not-yet-answered) friend requests. Specifically, \textsc{PreAttacK} updates the probability that a new user is fake to the extent she ($1$) `preferentially attaches' to specific recipients in keeping with their probabilities of being requested by fake accounts vs. by real ones, and also ($2$) to the extent existing users `preferentially attach' to her in keeping with their probabilities of sending requests to fake accounts vs. real ones.

\begin{itemize}
    \item \textbf{Theoretic contribution.} We derive instance-specific bounds that show \textsc{PreAttacK} near-optimally approximates each new user's posterior probability of being fake in relevant problem instances at lower computational cost than alternatives. These are the first provable guarantees for fake account detection that apply to new users, and that do not require strong homophily assumptions. Indeed, despite the enormous popularity of Preferential Attachment models, to our knowledge \textsc{PreAttacK} is the first time that the corresponding classifier has been derived.

\vspace{2mm}
\item \textbf{Real-world effectiveness.} This principled approach makes \textsc{PreAttacK} the only algorithm with provable guarantees that obtains state-of-the-art performance at scale on the global Facebook network. Specifically, we implement \textsc{PreAttacK} at scale at Facebook and show it obtains high AUC$\approx$$0.9$ after new users sent/received a total of just $20$ not-yet-answered friend requests. 
For comparison, state-of-the-art network-based algorithms do not obtain this performance even after observing additional data on new users' first $100$ friend requests.
This means that unlike existing algorithms, \textsc{PreAttacK} converges to detect fakes before the median new fake account makes a single friendship (i.e. \emph{accepted} request) with a real user (see Figs. \ref{fig:motivation_plots1} and \ref{fig:motivation_plots2} on the previous page).
\vspace{2mm}
\item \textbf{General applicability.} While we focus on fake accounts on Facebook, \textsc{PreAttacK} applies generally to networks where directed edges convey information about users' latent labels, such as Twitter or Instagram `follows',  LinkedIn `connects', etc. \textsc{PreAttacK} may also be used to infer new users' other latent class labels beyond fake/real (e.g. political party, etc.), which offers a means to address other cold-start problems.
\end{itemize}

\paragraph{\textbf{Paper organization.}}
Section \ref{sec:kCDPA} specifies the $k$CDPA model. Sections \ref{sec:algorithm} and \ref{sec:analysis} derive \textsc{PreAttacK} and its instance-specific approximation bounds. Section \ref{sec:PreAttacKpp} extends \textsc{PreAttacK} by incorporating observed homophily to obtain faster convergence. Section \ref{sec:eval} shows \textsc{PreAttacK}'s performance on the Facebook network.

\vspace{3mm}

\section{Multi-Class Directed PA ($k$CDPA)}
\label{sec:kCDPA}

Our core generative model is a simple but powerful extension of the canonical directed Preferential Attachment (\emph{rich-get-richer}) model to the setting where there are $k$=$2$ classes of new users, \emph{fakes} and \emph{reals}, who join a preexisting social network. Whereas traditional PA models capture how new users tend to seek out already-popular users, \hyperref[kCDPA]{$k$CDPA} captures how new users tend to seek out (request and/or be requested by) users who are already popular with those of the new user's fake/real class:

\begin{itemize}
    \item \textbf{New users' outgoing friend requests:} We model that new \emph{fake} users send friend requests to existing users drawn in proportion to the counts of requests existing users already received \emph{from fakes only}, and new \emph{real} users send friend requests to existing users drawn in proportion to the counts of requests existing users already received \emph{from reals only}.
    \item \textbf{New users' incoming friend requests:} Similarly, each new \emph{fake} [or \emph{real}] user \emph{receives} requests from existing users who are drawn in proportion to the counts of requests existing users already sent \emph{to fakes} [or \emph{reals}]. 
\end{itemize}

The $k$CDPA model is formally described by the following generative process. Suppose we have a preexisting directed social network $G(V,E_0,L_V)$ where edges $E_0$ capture friend \emph{requests} (\emph{not} friendships/accepted requests). We consider $k$$=$$2$ classes: users $\nu\in V$ have known fake/real labels $L_V:L_V\in \{F, R\}^{|V|}$. We will denote a single user $\nu$'s label by lowercase $\ell_\nu$. Finally, we have a small set of \emph{new users} $U:U=\{u_1\dots u_m\}$ who are each fake with probability $\pi$. 
Some new users $u$ are more likely than others to send a friend request and/or receive a friend request. To be as general as possible, suppose we have some distribution $\mathcal{D}$ that captures these probabilities (so $\mathcal{D}$'s domain includes $2|U|$ entries---two for the probability that each new user $u$ will $[send, receive]$ a friend request). The $k$CDPA model is then:



\begin{algorithm}[H]
\caption*{\textbf{$k$-Class Directed Preferential Attachment ($k$CDPA)}}
\addcontentsline{loa}{algorithm}{$k$-Class Directed Preferential Attachment}

\begin{algorithmic}
    	\INPUT Preexisting network of requests $G(V,E_0,L_V)$; \ new users $U$\\
    	\STATE Draw new users' fake/real labels $L_U = \{\ell_u \sim Bernoulli(\pi)\}$\\ 
    	\STATE \textbf{for} $i \in 1, \dots, n$\\
    	\STATE \ \ \ \ Draw new user \& direction $\{u\in U, \ d\in [send, receive]\} \sim \mathcal{D}$

    	\STATE \ \ \ \ \textbf{if} $d=send$\\
        \STATE   \ \ \ \ \ \ \ \ Draw $\nu\in V$; \  $P(\nu) \propto \alpha + \sum_{e_{x\rightarrow y} \in E_{i-1}} \mathbbm{1}[(y=\nu) \wedge (\ell_x=\ell_u)]$
        \STATE  \ \ \ \ \ \ \ \ $E_{i} = E_{i-1}\cup \{u \rightarrow \nu\}$\\

        \STATE \ \ \ \ \textbf{else} 
        \STATE   \ \ \ \ \ \ \ \ Draw $\nu\in V$; \ $P(\nu) \propto \alpha + \sum_{e_{x\rightarrow y} \in E_{i-1}} \mathbbm{1}[(x=\nu) \wedge (\ell_y=\ell_u)]$

        \STATE  \ \ \ \ \ \ \ \ $E_{i} = E_{i-1}\cup \{\nu \rightarrow u\}$\\
        \STATE \textbf{return} $G(V\cup U, \  E_i, \ L_V\cup L_U)$
  \end{algorithmic}
  \label{kCDPA}
\end{algorithm}

Here, $\alpha$ is a small constant that captures e.g. the probability that a preexisting user $\nu$ receives or sends her first-ever request (in Section \ref{sec:PreAttacKpp} below, we consider a `homophily-incorporating' extension where $\alpha$ depends on the sender and receivers' real/fake labels). By $\mathbbm 1$ we denote the indicator function that takes value $1$ if the argument is true and $0$ otherwise, so the sum under the \emph{if $d=send$} statement counts the number of friend requests that existing user $\nu$ has already received from users who have the same $\{fake,real\}$ label as new user $u$. Note that this includes requests from the preexisting network ($E_0$) as well as requests from new users in previous iterations.\footnote{As in the original PA model, this generative process may result in a multigraph (e.g. if the same edge is drawn twice). This is suitable for our setting, as social network users can send multiple friend requests to the same recipient (e.g. if the first is rejected).}

While we are interested in $k$$=$$2$ classes, note that \hyperref[kCDPA]{$k$CDPA} easily extends to the case where there are $k$$>$$2$ classes of users, $L_V \in \{1,\dots,k\}^{|V|}$ just by replacing \emph{Bernoulli($\pi$)} with \emph{Multinom($\pi_1,\dots, \pi_k)$}. This captures (for example) settings where there are multiple types of fake users: sockpuppets, false news bots \citep{vosoughi2018spread}, etc., and each has different preferences in terms of existing users they seek to befriend.

Very recently, similar 2-class (and multi-class) PA models have received much attention due to their ability to explain the generative process by which homophily and related properties emerge in social networks~\citep{pmlr-v38-lee15b, avin2015homophily, zhang2018understanding, nettasinghe2021directed, avin2020mixed}. However, for our purposes, we do not require the model to explain the full evolution of a social network; we merely require it to capture the friend request behavior of new users who join a long-established network (e.g. Facebook).

\section{The \textsc{PreAttacK} Algorithm}
\label{sec:algorithm}

In this section, we derive a new algorithm, \textbf{Pre}ferential \textbf{Attac}hment $\mathbf{k}$-class Classifier (\textsc{PreAttacK}) that near-optimally approximates the posterior probability that each new social network user is a fake account under the $k$CDPA model.
Intuitively, \textsc{PreAttacK} updates the probability that a new user is fake to the extent she (1) `preferentially attached' to specific recipients in keeping with their probabilities of being requested by fake accounts vs. by reals, and also (2) to the extent existing users `preferentially attached' to her in keeping with their probabilities of sending requests to fake accounts vs. reals. Because $k$CDPA models friend requests rather than friendships (accepted requests), \textsc{PreAttacK} can classify new users even before they make a single friendship. Surprisingly, despite the complex properties of $k$CDPA (and PA processes in general), we show that \textsc{PreAttacK} is also computationally efficient on mature social networks containing billions of users.

\paragraph{\textbf{\textsc{PreAttacK} considerations}} We are interested in the $k$$=$$2$ case where users are $\{real, fake\}$, but also show in Appendix \ref{appendix:multiclass} that \textsc{PreAttacK} also accommodates $k$$>$$2$ to classify multiple types of fakes, such as sockpuppets and false news bots.
Importantly, we will assume that the count of requests that each new user $u$ sends and receives are independent of her label. This precludes the undesirable scenario where \textsc{PreAttacK} e.g. penalizes new real users who send many requests by increasing the posterior probability that they are fake.
Finally, note that $k$CDPA generates no requests between new accounts. It is easy to modify $k$CDPA to generate such requests\footnote{To make this change, add a line: $V=V\cup u$ at the end of the for-loop.}, but excluding them precludes a scenario where the posterior probability that one new account is fake depends \emph{only} on other new accounts. This prevents malicious adversaries from manipulating \textsc{PreAttacK} by generating many new accounts at once (see Section \ref{ssec:adversarial}).


\noindent\paragraph{\textbf{\textsc{PreAttacK} part I: A new user's outgoing friend requests}}
The conditional probability $P_{\nu_i+}^F$ that new \emph{fake} user $u$ who \emph{sends} a friend request at iteration $i$ of \hyperref[kCDPA]{$k$CDPA} draws preexisting user $\nu$ for the recipient is proportional to the count of requests that $\nu$ already received from fakes before iteration $i$:
\begin{align}
P_{\nu_i+}^F \coloneqq& P[\nu_i|\mathbf{\ell_u=F}, \ E_{i-1}, u, d, L_V\cup L_{U}]\\
 =& \frac{ \alpha + \sum_{e_{x\rightarrow y} \in E_{i-1}}  \mathbbm{1}[y \! = \! \nu \wedge \ell_x \! = \! F] }{
\alpha|V|+
\sum_{e_{x\rightarrow y} \in E_{i-1}}  \mathbbm{1}[ \ell_x \! = \! F]    }
\label{Pedgefake_send}
\end{align}
 Similarly, if new user $u$ is real, this probability becomes:
 \begin{align}
P_{\nu_i+}^R \coloneqq P[\nu_i|\ell_u \! = \! R,\cdot] = \frac{ \alpha + \sum_{e_{x\rightarrow y} \in E_{i-1}}  \mathbbm{1}[y \! = \! \nu \wedge \ell_x \! = \! R] }{
\alpha|V|+
\sum_{e_{x\rightarrow y} \in E_{i-1}}  \mathbbm{1}[ \ell_x \! = \! R]  }
\label{Pedgereal_send} 
\end{align}

Given all new users' $\{F, R\}$ labels and the sequence of all other new users' friend requests $E_1,\dots$, then the joint conditional probability of observing $u$'s \emph{sequence} of outgoing friend request recipients $\nu_i$ is just  the product of their individual probabilities (eqn. \ref{Pedgefake_send} or \ref{Pedgereal_send}). Denote this sequence of $u$'s recipients by $\mathcal{N}_u^+$. If $u$ is fake:

\begin{align}
P_{\mathcal{N}_u^+}^F \coloneqq P[\mathcal{N}_u^+|\ell_u \! = \! F,\cdot] =   \prod_{\nu_i:\{u\rightarrow \nu\}_i\in \mathcal{N}_u^+}  P_{\nu_i+}^F
\label{Prodfake_send}
\end{align}

\noindent And similarly, if $u$ is real, this conditional probability is:
\begin{align}
P_{\mathcal{N}_u^+}^R \coloneqq P[\mathcal{N}_u^+|\ell_u \! = \! R,\cdot] =   \prod_{\nu_i:\{u\rightarrow \nu\}_i\in  \mathcal{N}_u^+}  P_{\nu_i+}^R
\label{Prodreal_send}
\end{align}

\paragraph{\textbf{\textsc{PreAttacK} part II: A new user's incoming requests}}
Noting the symmetry of the $k$CDPA model with respect to requests that new users send and receive, we can also derive the cond. probability $P_{\nu_i-}^F$ that a new user $u$ who \emph{receives} a friend request at iteration $i$ draws preexisting user $\nu$ for the request's sender. Similar to above, this probability is proportional to the count of requests that $\nu$ has already sent to users who share the same label as $u$. If $u$ is fake:

\begin{align}
P_{\nu_i-}^F \coloneqq P[\nu_i|\ell_u \! = \! F,\cdot] = \frac{\alpha + \sum_{e_{x\rightarrow y} \in E_{i-1}}  \mathbbm{1}[x \! = \! \nu \wedge \ell_y \! = \! F] }{
\alpha|V|+
\sum_{e_{x\rightarrow y} \in E_{i-1}}  \mathbbm{1}[ \ell_y \! = \! F]    }
\label{Pedgefake_receive}
\end{align}
And if new user $u$ is real, this conditional probability is:
 \begin{align}
P_{\nu_i-}^R \coloneqq P[\nu_i|\ell_u \! = \! R,\cdot] = \frac{\alpha + \sum_{e_{x\rightarrow y} \in E_{i-1}}  \mathbbm{1}[x \! = \! \nu \wedge \ell_y \! = \! R] }{
\alpha|V|+
\sum_{e_{x\rightarrow y} \in E_{i-1}}  \mathbbm{1}[ \ell_y \! = \! R]  }
\label{Pedgereal_receive} 
\end{align}

\noindent Similar to above, the joint conditional probability of $u$'s sequence of incoming friend request senders (denoted by $\mathcal{N}_u^-)$ if $u$ is fake is:

\begin{align}
P_{\mathcal{N}_u^-}^F \coloneqq P[\mathcal{N}_u^-|\ell_u \! = \! F,\cdot] =   \prod_{\nu_i:\{\nu\rightarrow u\}_i\in \mathcal{N}_u^-}  P_{\nu_i-}^F
\label{Prodfake_receive}
\end{align}

\noindent And similarly, if $u$ is real, this conditional probability is:
\begin{align}
P_{\mathcal{N}_u^-}^R \coloneqq P[\mathcal{N}_u^-|\ell_u \! = \! R,\cdot] =   \prod_{\nu_i:\{\nu\rightarrow u\}_i\in \mathcal{N}_u^-}  P_{\nu_i-}^R
\label{Prodreal_receive}
\end{align}

\paragraph{\textbf{Posterior probability that a new user is fake}} We are now able to derive the full posterior probability that new user $u$ is fake as a function of the observed sequence of preexisting users to whom she sent friend requests and from whom she received requests. Leveraging Bayes' rule and the law of total probability we have:

\begin{align}
\mathbf{P^*_u} :&= P[\ell_u=F | \mathcal{N}_u^+,\mathcal{N}_u^-, E_0,E_1\dots E_n, L_V\cup L_{U\backslash u}] \\&=
\frac{
P_{\mathcal{N}_u^+}^F \cdot P_{\mathcal{N}_u^-}^F \cdot \pi
}{
P_{\mathcal{N}_u^+}^F \cdot P_{\mathcal{N}_u^-}^F \cdot \pi \ + \ P_{\mathcal{N}_u^+}^R \cdot P_{\mathcal{N}_u^-}^R \cdot (1-\pi)
}\\&=
\Big( 1\ + \ (P_{\mathcal{N}_u^+}^F \cdot P_{\mathcal{N}_u^-}^F )^{-1}(P_{\mathcal{N}_u^+}^R \cdot P_{\mathcal{N}_u^-}^R) \cdot \pi^{-1}(1-\pi)\Big)^{-1}
\label{posterior}
\end{align}

This posterior captures the idea that $u$ is relatively more likely to be fake to the extent she `preferentially' sent requests to recipients who are more preferred by fakes, and also to the extent she received requests from senders who are more likely to send to fakes.

\subsection{Intractability}

Unfortunately, this expression for the posterior probability $\mathbf{P^*_u}$ that a new user $u$ is fake is intractable, as it requires knowledge of the (latent) real/fake label of all new users who sent requests before $u$. Moreover, computing this posterior in expectation becomes infeasible as we consider more than a handful of new users, as this requires integrating over all possible label combinations.

A standard approach at this point would be to apply either linearized belief propagation or MCMC techniques. However, both are computationally expensive in large networks due to the need to e.g. iterate between inferring new users' posterior labels and updating \emph{all} existing users' sending and receiving preferential attachment weights (i.e. sums within $P_{\nu_i+}^F, P_{\nu_i+}^R, P_{\nu_i-}^F, P_{\nu_i-}^R$) until (possible) convergence. They also typically lack convergence guarantees \citep{yedidia2003understanding, gong2014sybilbelief}, or obtain guarantees only at the expense of the approximation (e.g. via linearization) \citep{wang2017sybilscar, wang2017gang} or significant complexity \citep{yedidia2000generalized}.

\subsection{\textbf{Fast approximation}}
\label{threeintuitions}
In contrast to these approaches, we consider a fast approximation for $\mathbf{P^*_u}$ based on the following idea: PA probabilities in mature social networks are stable over small batches of new entrants. \emph{So, rather than account for small and intractable changes to one new user's posterior that accrue due to other new users' edges $E_1,\dots$, we ignore them and then bound their worst-case impact.}  Consider that given a large preexisting network, a small batch of new accounts who send and receive friend requests (probably) do \emph{not} significantly change existing users' PA probabilities (i.e. sums in the \emph{Draw} steps of \hyperref[kCDPA]{$k$CDPA}). At a high level, there are three reasons why this is so:

\begin{enumerate}
    \item \textbf{Collisions are (probably) rare.} Given a large preexisting network of $300$ million (Twitter) or $2$ billion (Facebook) users, a small batch of new users are unlikely to `draw' the same recipients multiple times. 
     When a new user sends a request to a recipient who was not previously requested by a new user, the numerators in $P_{\nu_i+}^F$ and $P_{\nu_i+}^R$ are equal to their (known) original values in $E_0$. The same is true of numerators in $P_{\nu_i-}^F$ and $P_{\nu_i-}^R$ when a new user receives a request from a not-previously-drawn sender.
    \item \textbf{Collisions (probably) have negligible impact.} In cases where multiple new accounts \emph{do} send friend requests to the same preexisting recipient, that recipient was probably already very popular (i.e. already had a large PA probability) due to PA's `rich-get-richer' dynamics. In that case, this preexisting recipient's PA probability only undergoes a small percentage change after each new request, so it is well-approximated by its original value in $E_0$. This argument also applies when multiple new accounts \emph{receive} requests from the same preexisting recipient.
    \item \textbf{New users have a small number of friend requests.} A large preexisting social network of billions of users results from on the order of $10^{11}$ friend requests. The new requests sent by a relatively small batch of new fake and real accounts has only a negligible impact on this preexisting count. Therefore, the denominators in $P_{\nu_i+}^F$, $P_{\nu_i+}^R$, $P_{\nu_i-}^F$, $P_{\nu_i-}^R$ are well-approximated by their original values in $E_0$.
\end{enumerate}

These three key intuitions, which we formalize in Section \ref{sec:analysis}, suggest  we can obtain a good approximation for the posterior $\mathbf{P^*_u}$ by holding all PA probabilities fixed at their values in the preexisting requests network $G(V,\mathbf{E_0},L_V)$. With this change, we can approximate the probability of observing the $i$'th request that a new (fake or real) account sends or receives, $P_{\nu_i+}^F, P_{\nu_i+}^R, P_{\nu_i-}^F$, and $P_{\nu_i-}^R$, \emph{without} knowing the labels of other new accounts. For example, for the `sending' probabilities $P_{\nu_i+}^F$ and $P_{\nu_i+}^R$:

\begin{align}
P_{\nu_i+}^F &= P[\nu_i|\ell_u=F,\cdot] \\ &\approx \hat{P}_{\nu+}^F \coloneqq \frac{\alpha + \sum_{e_{x\rightarrow y} \in \mathbf{E_0}}  \mathbbm{1}[y=\nu \wedge \ell_x=F] }{
\alpha|V| + \sum_{e_{x\rightarrow y} \in \mathbf{E_0}} \mathbbm{1}[ \ell_x=F]    }
\label{Pedgefake_send_APPROX}
\end{align}
And similarly:
\begin{align}
P_{\nu_i+}^R &= P[\nu_i|\ell_u=R,\cdot] \\ &\approx \hat{P}_{\nu+}^R \coloneqq \frac{\alpha + \sum_{e_{x\rightarrow y} \in \mathbf{E_0}}  \mathbbm{1}[y=\nu \wedge \ell_x=R] }{
\alpha|V| + \sum_{e_{x\rightarrow y} \in \mathbf{E_0}} \mathbbm{1}[ \ell_x=R]    }
\label{Pedge_real_send_APPROX}
\end{align}

We obtain approximations for the remaining PA probabilities (`receiving' probabilities) $\hat{P}_{\nu-}^F$, and $\hat{P}_{\nu-}^R$ by making the identical substitution of $E_0$ for $E_{i-1}$ in eqns. \ref{Pedgefake_receive}, and \ref{Pedgereal_receive} (note that these four approximations are constant for all new edges to/from the same preexisting user $\nu$, so we drop $i$ subscripts accordingly). 

We now obtain an approximation $\mathbf{\hat{P}_u}$ of the posterior probability $\mathbf{P^*_u}$ that new user $u$ is fake by using these approximations in eqns. \ref{Prodfake_send}, \ref{Prodreal_send}, \ref{Prodfake_receive}, and \ref{Prodreal_receive} to approximate the joint probabilities of all of user $u$'s outgoing \& incoming edges conditional on her real/fake label, $\hat{P}_{\mathcal{N}_u^+}^F, \  \hat{P}_{\mathcal{N}_u^+}^R, \ \hat{P}_{\mathcal{N}_u^-}^F \ $, and $\hat{P}_{\mathcal{N}_u^-}^R$,  then computing her posterior (eqn. \ref{posterior}). This approach is formalized in the \textsc{PreAttacK} algorithm:


\begin{algorithm}[H]
\caption*{\textbf{PreAttacK}}
\addcontentsline{loa}{algorithm}{PreAttacK}

\begin{algorithmic}
    	\INPUT Preexisting $G(V,E_0,L_V)$; \ new users $U$; new requests $E_n\backslash E_0$ \\
    	\STATE \textbf{for} $\nu \in V$ who receives a new request, $\bigcup (\nu: \{x\rightarrow \nu\} \in E_n\backslash E_0$)\\
        \STATE \ \ \ \ Compute $\hat{P}_{\nu+}^F$ and $\hat{P}_{\nu+}^R$   \\
    	\STATE \textbf{for} $\nu\in V$ who sends a new request, $\bigcup (\nu: \{\nu\rightarrow x\} \in  E_n\backslash E_0$)\\
    	\STATE \ \ \ \ Compute $\hat{P}_{\nu-}^F$ and $\hat{P}_{\nu-}^R$ \\
    	\STATE \textbf{for} new user $u\in U$\\
    	\STATE \ \ \ \ Compute $\hat{P}_{N_u^+}^F, \  \hat{P}_{N_u^+}^R, \ \hat{P}_{N_u^-}^F, \ $ and $\hat{P}_{N_u^-}^R$\\
    	\STATE \ \ \ \ Compute posterior $\mathbf{\hat{P}_u}$     
        \STATE \textbf{return} $[\mathbf{\hat{P}_1},\dots, \mathbf{\hat{P}_{|U|}}]$
  \end{algorithmic}
  \label{alg:PreAttacK}
\end{algorithm}


Below, we show that in our setting \textsc{PreAttacK} obtains near-optimal approximations for the posterior probabilities $\mathbf{P^*_u}$ at low computational cost. We also show in Section \ref{sec:PreAttacKpp} that it can be naturally extended to capture \emph{homophily} or even \emph{monophily}---scenarios where $\alpha=\function(\ell_\nu, \ell_u)$. These extensions incur no cost in terms of complexity, and they slightly improve the approximation bounds.

\section{Analysis of \textsc{PreAttacK}}
Our goal in this section is to show that \textsc{PreAttacK} results in improved computational complexity over alternatives, and that it admits instance-specific approximation bounds that confirm near-optimal posterior inference for our problem instance. We note that these are some of the first theoretic guarantees for this problem that do not rely on homophily assumptions.
\label{sec:analysis}

\subsection{\textbf{Complexity of \textsc{PreAttacK}}}
Computing all existing users' preferential attachment weights requires $|E_0| + 4(|V^+| + |V^-|) \le |E_0| + 8(|V|)$ simple operations, where $V^+, V^- \subseteq V$ respectively refer to the subset of preexisting users who receive and send requests in preexisting network $E_0$. Then, computing \textsc{PreAttacK}'s posterior for all new accounts $U$ requires $2|E_n\backslash E_0| + 2|U|$ operations. Importantly, unlike state-of-the-art algorithms, \textsc{PreAttacK} can be computed for all new accounts in a single pass through all edges \citep{gong2014sybilbelief, yang2012sybilrank, wang2017sybilscar, wang2017gang}. 
This yields $\mathcal{O}(|E_n|)$ asymptotic complexity, which is $\mathcal{O}(|V\cup U|)$ in (sparse) social networks \citep{mislove2007measurement}. This improves on state-of-the-art algorithms such as \textsc{SybilBelief}, \textsc{SybilRank}, and \textsc{SybilSCAR}, which require $\mathcal{O}(m|E'|)$, where $m$ is the number of iterations (at least $\mathcal{O}(\log(|V\cup U|))$) and $E'$ is the set of all accepted friend requests~\citep{gong2014sybilbelief, yang2012sybilrank, wang2017sybilscar}.

\subsection{\textbf{Instance-specific approximation guarantee}}
We formalize the three key intuitions from Section \ref{threeintuitions} to derive instance-specific and new-user-specific approximation guarantees. This is advantageous because it allows researchers to also obtain an upper- and lower-bound of the \emph{exact} posterior for each new user, and also to determine the batch size (or subset) of new users that can be classified while maintaining a desired worst-case approximation bound for a specific problem instance. We give the key intuition for the proof here and defer full analysis to Appendix \ref{boundproof}.

\paragraph{\textbf{One-sided approximation errors}} It is acceptable for \textsc{PreAttacK} to overestimate the posterior probability that a new fake is fake and underestimate the probability that a new real is fake, but not the opposite. 
Therefore we seek, for each new user $u$, two bounds: a worst-case approximation factor (underestimate factor) $f^F \le \mathbf{\hat{P}_u}/\mathbf{P^*_u}$, which is useful if $u$ is fake, and a factor (overestimate factor) $f^R \ge \mathbf{\hat{P}_u}/\mathbf{P^*_u}$ that is useful in case $u$ is real. 

\paragraph{\textbf{Avoiding the combinatorial problem of new users' labels}}
Consider $f^F$. 
The main difficulty is that we cannot know (without trying all combinations) the worst-case configuration of new users' latent labels that results in the worst underestimate $\mathbf{\hat{P}_u}/\mathbf{P^*_u}$. This is because each new user before $u$ may have sent multiple requests to recipients $\nu$, some of which result in increases to $\mathbf{P^*_u}$ (e.g. if the other new user is also fake and targets some of the same recipients as $u$) and some in decreases (e.g. if the other new user is also fake and targets some recipients who are not among $u$'s recipients).

We sidestep this combinatorial problem by imagining that each new edge to/from a new account prior to $u$'s is sent by a \emph{unique} `phantom' new account $p$ whose label is the worst-case label for the bound of interest. Thus, for $f^F$ we assume $\ell_p$=$F$ if $p$'s single new request is to/from the same preexisting recipient $\nu$ as one of $u$'s requests, and $\ell_p$=$R$ otherwise. 
Compute $u$'s `worst case underestimate if $u$ is fake' posterior $\mathbf{P^F_{u, WC}}$ using these `phantom labels' to obtain $f^F = \mathbf{\hat{P}_u}/\mathbf{P^F_{u, WC}} \le \mathbf{\hat{P}_u}/\mathbf{P^*_u}$. To then obtain the `worst case overestimate if $u$ is real' factor $f^R$, compute  $\mathbf{P^R_{u, WC}}$ assuming the opposite: $\ell_p$=$R$ if $p$'s single new request is to/from the same preexisting user $\nu$ as one of $u$'s requests, else $\ell_p$=$F$ (see Appendix \ref{boundproof}).

In Section \ref{sec:eval}, we show this yields useful approximation bounds for millions of new accounts in real data ($f^F\approx 0.85, f^R\approx 1.1$).

\subsection{Adversarial robustness in practice} \label{ssec:adversarial}
We also highlight an important property that \textsc{PreAttacK} shares with recent advances in practical adversarial robustness for this problem. The most performant recent algorithms for fake account detection at Facebook obtain adversarial robustness in practice by leveraging so-called `deep network features' \citep{xudeep}, which are features that capture aggregate properties of each user's friends-of-friends. Such aggregates have been shown to be practically difficult for even coordinated campaigns of fake accounts to manipulate, particularly when befriending (at least some) real users. \textsc{PreAttacK} similarly works by aggregating over the features (i.e. counts) of e.g. friend-requesters-of-friend-requestees. As such, \textsc{PreAttacK}'s preferential attachment probabilities may also be considered `deep network features'. Manipulating \textsc{PreAttacK}'s prediction for a certain user would require an adversary to manipulate the counts of fake and real senders who send requests to the user's recipients, as well as the counts of known fake and real users to whom the user's requesters also send requests.\footnote{Alternatively, a sophisticated adversary might attempt to learn and then target the set of real users who are primarily targeted by real users and not fakes (i.e. who have small $\hat{P}_{\nu+}^F / \hat{P}_{\nu+}^R <1$). However, even if this were possible, selection bias dictates that these real users may be less receptive to accepting fakes' friend requests, and the adversary would have to severely limit its fake accounts' friend requests to each real user $v$ to avoid increasing $\hat{P}_{\nu+}^F$ (which would result in future detection by \textsc{PreAttacK}).} See also Appendix \ref{appendix:selectionbias}.

Below, we also consider a variant of \textsc{PreAttacK} called \textsc{PreAttacK++} that also prevents sophisticated adversaries from avoiding detection by targeting only very unpopular (and thus uninformative) real users who have sent and received few friend requests.

Finally, we note that in practice on large scale social networks, new approaches to this problem that are practically vulnerable to attack (such as modifying a fake account classifier by adding a new and informative feature that can be manipulated by users) tend to prompt an observable response from sophisticated adversaries (see e.g. \citep{xudeep}). We have observed no such response to \textsc{PreAttacK}.

\section{\textsc{PreAttacK++}  and  Homophily}
\label{sec:PreAttacKpp}

We also consider a variant of  \textsc{PreAttacK}, \textsc{PreAttacK++}, that incorporates \emph{homophily} and/or \emph{monophily}\footnote{Recall that monophily occurs where one type of user prefers to connect to a specific other type of user, e.g. if fake users send requests to reals rather than other fakes.} to more rapidly detect fakes. \textsc{PreAttacK++} captures scenarios where the \hyperref[kCDPA]{$k$CDPA} prior probabilities\footnote{We refer to $\alpha$'s as `probabilities' for readability, but note that in \hyperref[kCDPA]{$k$CDPA}, PA probabilities are \emph{proportional} to $\alpha$, so it is possible to choose parameters $\alpha \in [0,\inf)$.} $\alpha$ that an existing user $\nu$ receives a request from (or sends a request to) a new user $u$ depend on $u$ and $\nu$'s real/fake labels, and also on whether the new account is the sender or the recipient. This captures e.g. a typical case where a new real account is \emph{a priori} much less likely to send a request to a preexisting fake account vs. a preexisting real account (even if neither has previously received any requests). It can also capture \emph{monophilic} networks where e.g. new fakes prefer to target real users rather than other fakes.

Incorporating these label-dependent probabilities is advantageous because they allow the posterior to update even when a new user sends requests to (or receives requests from) preexisting recipients who have not received any requests, but whose label is known. This also prevents sophisticated fake accounts from avoiding detection by targeting only unpopular recipients.

In the most general case, $\alpha$ can take $8$ values: $4$ probabilities that a new $\{fake,real\}$ user $u$ \emph{sends} a request to any preexisting $\{fake,real\}$ user $\nu$, which we denote by $\alpha_{\ell_u\rightarrow\ell_\nu}^+$ and $4$ probabilities that a new $\{fake,real\}$ user $u$ \emph{receives} a request from any preexisting $\{fake,real\}$ $\nu$, denoted by $\alpha_{\ell_\nu\rightarrow\ell_u}^-$. Estimates of these probabilities are known or easily obtainable from historical data. \textsc{PreAttacK++} uses them (per \hyperref[kCDPA]{$k$CDPA}) in the approximate probabilities $\hat{P}_{\nu+}^F$, $\hat{P}_{\nu+}^R$, $\hat{P}_{\nu-}^F$, and $\hat{P}_{\nu-}^R$ of observing each new edge in the first $2$ loops in \textsc{PreAttacK}. For example, in \textsc{PreAttacK++}, the probability $\hat{P}_{\nu+}^F$ that a new fake user sends a request to $\nu$ becomes:

\begin{align}
 \hat{P}_{\nu+}^F = \frac{\mathbf{\alpha_{F\rightarrow\ell_\nu}^+} +  \sum_{e_{x\rightarrow y} \in \mathbf{E_0}}  \mathbbm{1}[y=\nu \wedge \ell_x=F] }{
\sum_{v\in V}  \mathbf{ \alpha_{F\rightarrow\ell_\nu}^+} \ + \ 
\sum_{e_{x\rightarrow y} \in \mathbf{E_0}}   \mathbbm{1}[ \ell_x=F] \big)   }
\label{Pedgefake_send_APPROX_PREATTACKPP}
\end{align}
And the probability a new fake receives a request from $\nu$ becomes:
\begin{align}
 \hat{P}_{\nu-}^F = \frac{\mathbf{\alpha_{\ell_\nu\rightarrow F}^-} +  \sum_{e_{x\rightarrow y} \in \mathbf{E_0}}  \mathbbm{1}[x=\nu \wedge \ell_y=F] }{
\sum_{v\in V}  \mathbf{ \alpha_{\ell_\nu\rightarrow F}^-} \ + \ 
\sum_{e_{x\rightarrow y} \in \mathbf{E_0}}   \mathbbm{1}[ \ell_y=F] \big)   }
\label{Pedgefake__recv_APPROX_PREATTACKPP}
\end{align}
Note that \textsc{PreAttacK++}'s new expressions for $\hat{P}_{\nu+}^R$ and  $\hat{P}_{\nu-}^R$ can be obtained by substituting $R$ for $F$ everywhere in eqns. \ref{Pedgefake_send_APPROX_PREATTACKPP} and \ref{Pedgefake__recv_APPROX_PREATTACKPP}.

Note this change does not incur a penalty in terms of complexity. Also, because more informative $\alpha_{\ell_u\rightarrow\ell_\nu}^+$ and $\alpha_{\ell_\nu\rightarrow\ell_u}^-$ values reduce the marginal change in posterior that can accrue due to new edges in each existing user's PA weights, \textsc{PreAttacK++} admits slightly improved instance-specific bounds compared to \textsc{PreAttacK} for identical problem instances (see Appendix \ref{boundproof}).

\section{Evaluations}
\label{sec:eval}

Our goal in this section is to show that beyond its provable guarantees, \textsc{PreAttacK} performs well in practice on new fake accounts on the global Facebook network. Our goal is \emph{not} to measure performance on \emph{all} fake accounts, as the current generation of production classifiers already detect the vast majority of fakes during account registration \citep{yang2012sybilrank, xudeep}. Similarly, \textsc{PreAttacK} is not an alternative to other production classifiers that detect longer-tenured fake accounts based on their longer timelines of friendships and shared content \citep{noorshams2020ties}. Rather, we seek to overcome the \emph{early detection paradox} by rapidly obtaining a good classification after an account passes registration, but before it can engage with real users.  Thus, rather than measure performance on all new accounts (including those easily detected by existing means), we instead evaluate the degree to which \textsc{PreAttacK} \emph{improves upon state-of-the-art defenses already in place} \citep{xudeep, breuer2020friend, noorshams2020ties} by detecting new fake accounts that are not yet detected by those methods. This `hardest-to-detect' class \citep{xudeep, kozlov2020evaluating, breuer2020friend} of new fakes motivates our evaluations.

Our main empirical result is that \textsc{PreAttacK} converges to informative classifications (AUC $\approx$$0.9$) after new accounts send + receive a total of $20$ not-yet-answered friend requests.\footnote{As is standard, we use the ROC AUC as our metric because real/fake account labels are highly imbalanced ($\sim$$95$$\%$ of users are real) \citep{xudeep, wang2018structure, breuer2020friend}. Recall that a perfect classifier has AUC=$1$, whereas AUC=$0.5$ denotes `no better than random'.} For comparison, state-of-the-art network-based algorithms do not obtain this performance even after observing additional data on new users' first $100$ friend requests. This means that unlike many state-of-the-art algorithms, \textsc{PreAttacK} converges before the median fake account makes a single friendship (accepted request) with a real user.

To accomplish this, we conduct two sets of evaluations. In the first set, we evaluate \textsc{PreAttacK} and its variants on new accounts that joined the global Facebook network, and we show how \textsc{PreAttacK} converges to AUC $\approx$$0.9$ as each new account sends and receives its first handful of friend requests. In our second set of evaluations, we compare \textsc{PreAttacK} to four state-of-the-art network-based benchmarks. Because these benchmarks are significantly more computationally intensive than \textsc{PreAttacK}, we restrict our data in this 2nd evaluation to a single country of $\sim$$1$ million users.

\vspace{3mm} 
\subsection{Evaluation $1$ framework}
To evaluate \textsc{PreAttacK}'s performance on new fake accounts on the global Facebook network, we adopt the evaluation framework of \citep{breuer2020friend}. Specifically, we consider the set of all ($n$$>$$10^6$) new accounts that joined the global Facebook network during a particular week last year, along with the time-ordered set of friend requests that they sent and received during that week. Our goal is to determine whether \textsc{PreAttacK} could have accurately classified these new accounts \emph{using just their initial $1,2,\dots50$ initial friend requests from this first week after they joined the network}, based on the counts of requests that preexisting accounts had sent and received from real and fake accounts prior to the start of this week (i.e. preexisting users' PA probabilities). Because several months have passed since this `historical evaluation week', we can now measure the accuracy of \textsc{PreAttacK}'s `early' classifications against high-confidence labels subsequently obtained from production classifiers. \citep{xudeep, kozlov2020evaluating}. 

We also confirm \textsc{PreAttacK} guarantees near-optimal approximations ($f^F \ge 0.85, f^R \le 1.1$) for $>$$90\%$ of these new accounts by computing the instance-specific bounds (see Section \ref{sec:analysis}).

\begin{figure}[t]
\centering
\includegraphics[height=0.3\textwidth]{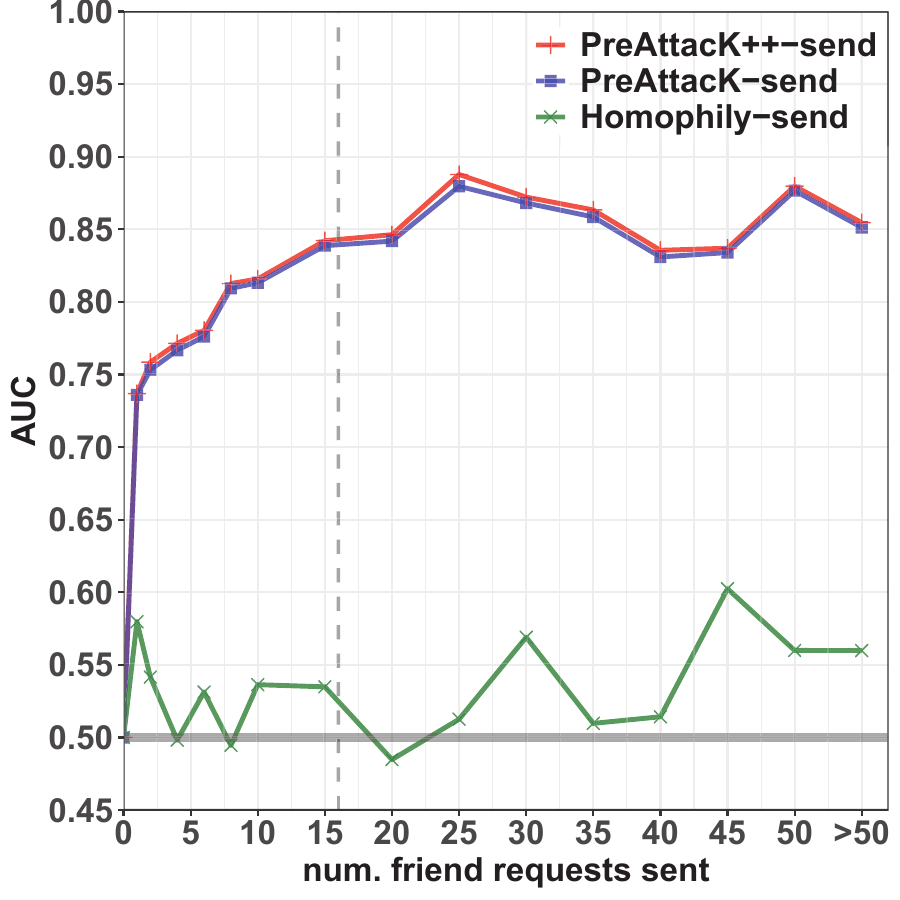}
\caption{Eval. 1 -\textsc{send} version AUC vs. \# friend requests sent.}
\Description[Performance on Global Facebook Network]{} 
\label{fig:RMglobal}
\vspace{-3mm}
\end{figure}

\paragraph{\textbf{\textsc{Homophily} benchmark}} We also consider a simplified variant of \textsc{PreAttacK}: \textsc{Homophily}. \textsc{Homophily} is identical to \textsc{PreAttacK++} but with existing users' PA probabilities zeroed out except for $\alpha$ terms, such that the probability of each new user's edge to/from \emph{any} existing user is proportional to the overall within- or cross-class rate $\alpha_{\ell_u\rightarrow\ell_\nu}^+$ or $\alpha_{\ell_\nu\rightarrow\ell_u}^-$ (see Appendix \ref{appendix:evaluations}). 
By comparing \textsc{PreAttacK} to \textsc{Homophily}, we ascertain the degree to which \textsc{PreAttacK}'s performance is homophily-based (i.e. driven by real vs. fake users' different preferences for in-class vs. cross-class friends) versus the degree to which it is driven by differences between real and fake users' preferences for individuals (i.e. our 2-Class PA model).

\paragraph{\textbf{\textsc{PreAttacK-send}, \textsc{PreAttacK++-send}, \& \textsc{Homophily-send}}} For each variant, we also compute a `\textsc{-send}' version that only considers the friend requests that new users \emph{sent} (and ignores requests they received). By comparing (for example) \textsc{PreAttacK-send} to \textsc{PreAttacK}, we measure how \textsc{PreAttacK}'s performance is driven by the requests that new users send vs. the requests they receive.

\paragraph{\textbf{Fast implementation and practical scaling}}
We implement \textsc{PreAttacK} and its variants in PyTorch \citep{NEURIPS2019_bdbca288}. On a 40-core 2GHz production virtual machine and even without GPUs, \textsc{PreAttacK} classifies more than a million new accounts-per-second. This efficiency permits us to recompute \textsc{PreAttacK}'s posterior after each user's first friend request, second request, and so on in order to obtain real-time-updated classifications for all new accounts.

\begin{figure}[t]
\centering
\includegraphics[height=0.3\textwidth]{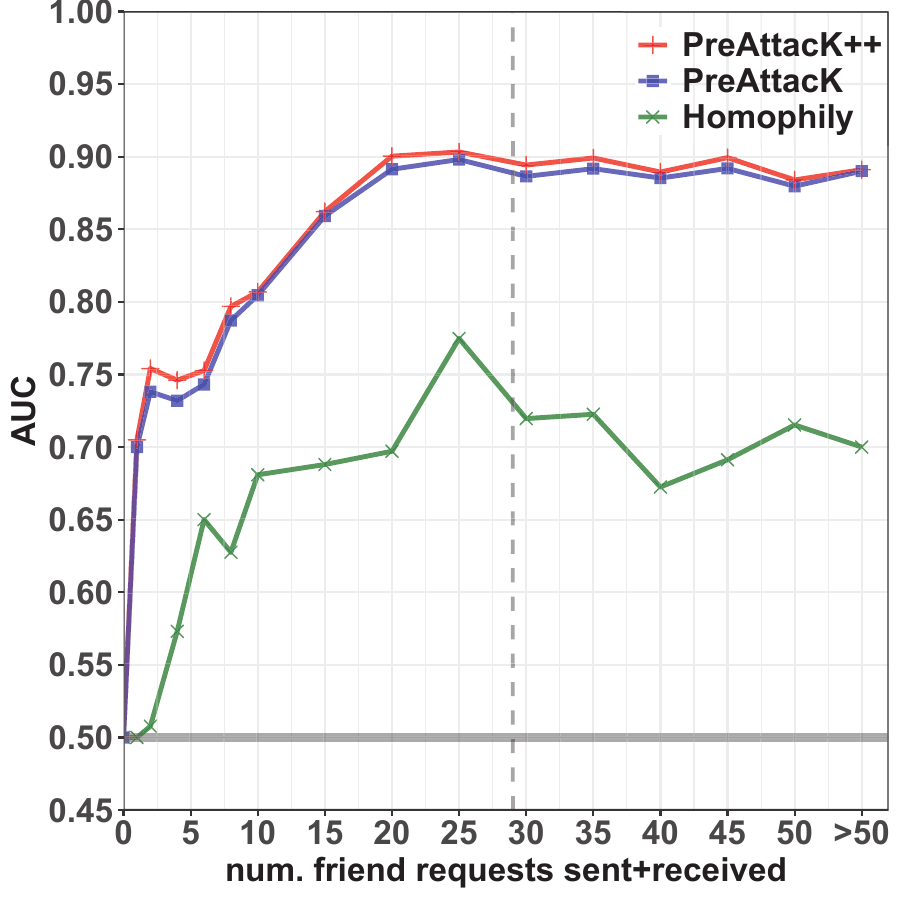}
\caption{Eval. 1 AUC vs. \# friend requests (sent+received).}
\Description[Performance on the Global Facebook Network]{} 
\label{fig:combglobal}
\vspace{2mm}
\end{figure}

\subsection{Evaluation $1$ Results}
Fig. \ref{fig:RMglobal} plots the AUC of \textsc{PreAttacK-send} versus the count of friend requests \emph{sent} by new accounts. Each $(x,y)$ point in the plot represents the AUC of the corresponding variant of \textsc{PreAttacK} run on just the first $x$ friend requests sent by new accounts during the `evaluation week'. Here, we observe that \textsc{PreAttacK-send} and \textsc{PreAttacK++-send} already obtain an informative posterior (AUC$>$0.75) after a new account sends $2$ friend requests---well less than the $16$ requests it takes the median new fake to make a friendship (i.e. \emph{accepted} request) with a real user. Note that the x-axis of Fig. \ref{fig:RMglobal} corresponds to our motivating plot, Fig. \ref{fig:motivation_plots1} in Section \ref{intro}. \textsc{PreAttacK-send} and \textsc{PreAttacK++-send} then converge to approx. AUC$\approx$$0.85$ after a new account sends $\approx$$25$ friend requests. 

Fig. \ref{fig:combglobal} plots the AUC of the full (send+receive) version of \textsc{PreAttacK} versus the total count of friend requests sent+received by new accounts. Here, the additional information regarding the friend requests that new accounts \emph{receive} permits \textsc{PreAttacK} and \textsc{PreAttacK++} to obtain AUC$\approx$$0.9$ after each new account sends + receives a total of 20 requests. Thus, they converge before the median fake account makes a friendship (i.e.\emph{accepted} request) with a single real user (which requires a \emph{total} of $29$ requests---see Fig. \ref{fig:motivation_plots2}). 

\bigskip
\paragraph{\textbf{\textsc{PreAttacK} vs. \textsc{Homophily}}}
Interestingly, \textsc{Homophily-send} performs only slightly better\footnote{This suggests that the `hardest-to-detect' new fake accounts in our evaluation set are savvy enough to avoid `suspicious' friendships with other fakes.} than random (Fig. \ref{fig:RMglobal}), and \textsc{Homophily} (Fig. \ref{fig:combglobal}) is only moderately informative. The large gap between \textsc{Homophily} vs. \textsc{PreAttacK} suggests that \textsc{PreAttacK}'s performance is driven by differences between real and fake users' preferences for individuals (i.e. \hyperref[kCDPA]{$k$CDPA}), rather than by real and fake users' different preferences for in-class vs. cross-class friends (i.e. homophily).


\paragraph{\textbf{\textsc{PreAttacK} vs. \textsc{PreAttacK++}}}
In both Fig. \ref{fig:RMglobal} and Fig. \ref{fig:combglobal}, \textsc{PreAttacK++} (or \textsc{PreAttacK++-send}) offers a small-but-consistent performance improvement of $\sim$$0.01$-$0.02$ AUC over \textsc{PreAttacK} (or \textsc{PreAttacK-send}), which is  considered nontrivial in this competitive domain \citep{xudeep, noorshams2020ties}. 
We compared them and found that `++' versions detected additional fakes that were targeting only `unpopular' existing users whose PA probabilities for both reals and fakes were both small (and thus less informative).

\subsection{Evaluation $2$ framework}
Evaluation 2 compares \textsc{PreAttacK} and its variants to four state-of-the-art network-based fake account detection algorithms: \textsc{GANG} \citep{wang2017gang}, \textsc{SybilRank}, \citep{yang2012sybilrank}, \textsc{SybilBelief} \citep{gong2014sybilbelief}, and \textsc{SybilSCAR} \citep{wang2018structure}.
\noindent These benchmarks are significantly more computationally intensive than \textsc{PreAttacK}, so we follow \citep{breuer2020friend} and restrict the network to a single country of $\sim$1 million users. This makes it practically feasible to run benchmarks using their papers' original C++ code and parameters. We provide details in Appendix \ref{appendix:benchmarks}.


\begin{figure}[t]
\centering
\includegraphics[height=0.3\textwidth]{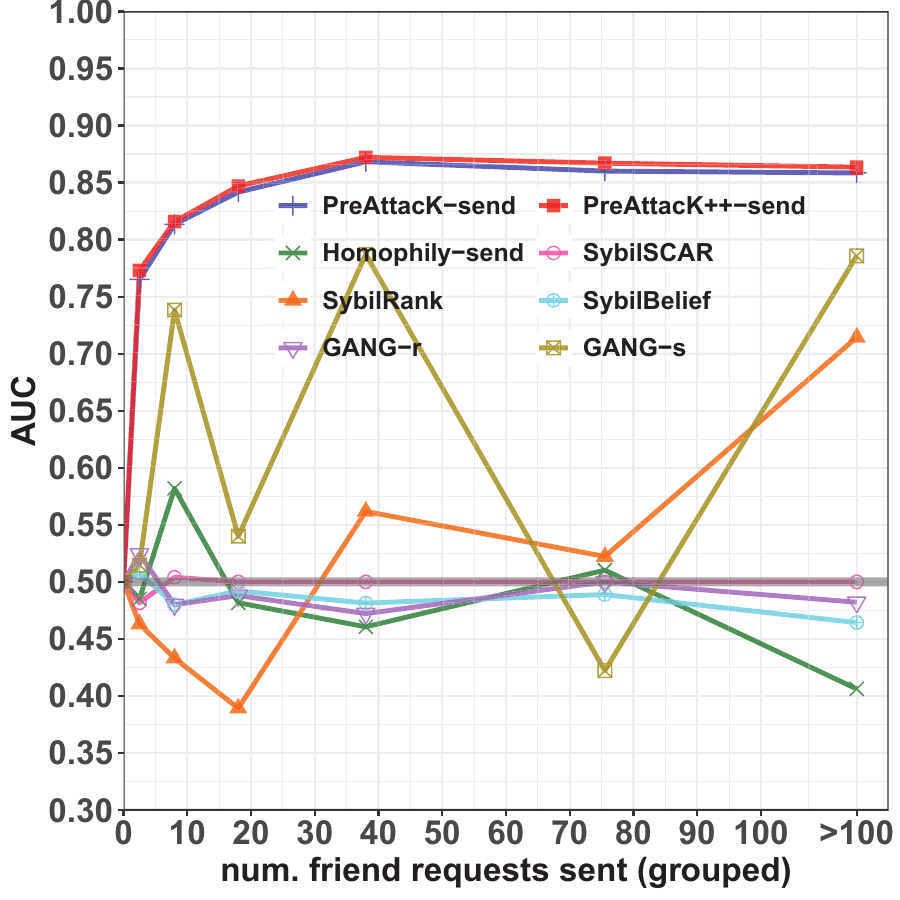}
\caption{Eval. 2 AUC vs. \# friend requests sent.}
\Description[Performance vs. benchmarks.]{} 
\label{fig:RMevalcountry}
\vspace{2mm}
\end{figure}

It is computationally impractical to run benchmarks multiple times to compute AUC after each new user's 1\textsuperscript{st}, 2\textsuperscript{nd}, etc. request, so we instead partition new users in Figs. \ref{fig:RMevalcountry} \& \ref{fig:combevalcountry} by \#requests they sent (or sent+received): $[0,5],[6,10],[11,25],[26,50],$ $[51,100],[101,\infty]$. 

\subsection{Evaluation $2$ Results}
\begin{figure}[t]
\centering
\includegraphics[height=0.3\textwidth]{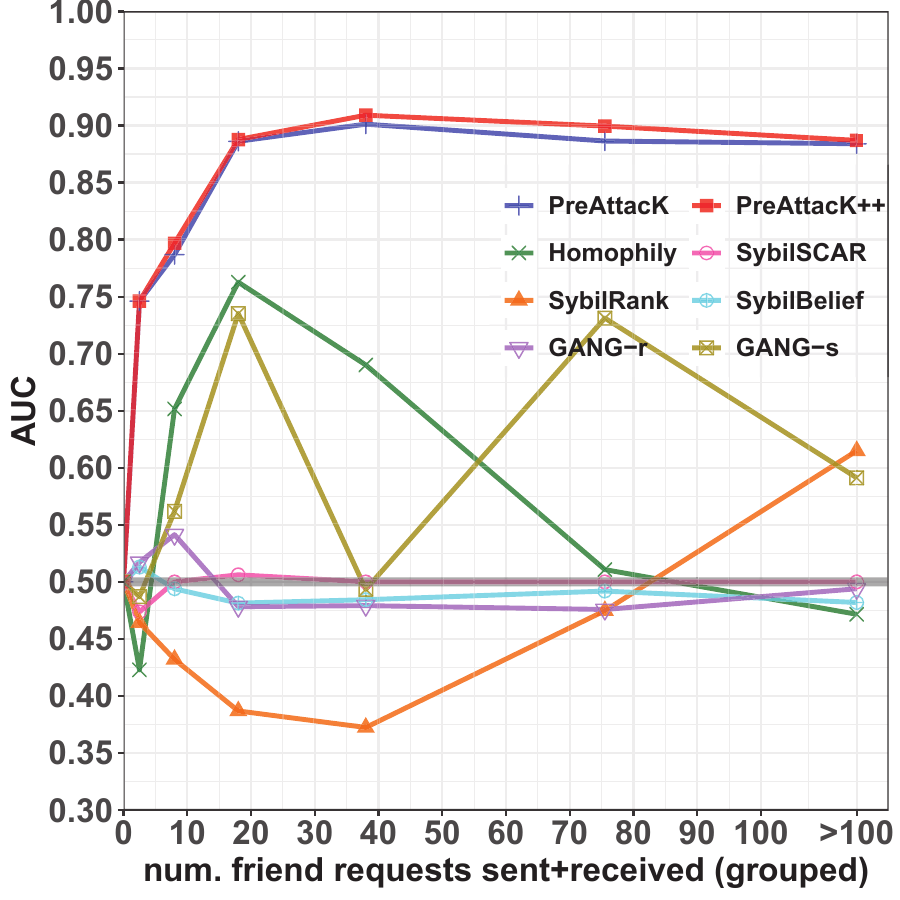}
\caption{Eval. 2 AUC vs. \# friend requests (sent+received).}
\Description[Performance vs. benchmarks.]{} 
\label{fig:combevalcountry}
\vspace{-3mm}
\end{figure}

Fig. \ref{fig:RMevalcountry} plots the AUC of \textsc{PreAttacK-send} and benchmarks vs. the count of friend requests that new accounts \emph{send}, and Fig. \ref{fig:combevalcountry} plots the AUC of full \textsc{PreAttacK} vs. the total count of requests that new accounts send+receive. Consistent with their performance on the global Facebook network (Figs. \ref{fig:RMglobal} \& \ref{fig:combglobal}), \textsc{PreAttacK-send} and \textsc{PreAttacK} obtain an informative signal of new accounts' authenticity before the median fake obtains a friendship (\emph{accepted} request) with a single human. In contrast, benchmarks perform poorly on new users, consistent with \citep{breuer2020friend}. We theorize this is because the current generation of new fakes do not exhibit sufficient homophily. \textsc{GANG-s} is a partial exception: it uses the directed network of friend requests (like \textsc{PreAttacK}) to obtain a useful AUC of $0.75$-$0.85$, albeit with high variance (Fig. \ref{fig:RMevalcountry}). However, unlike \textsc{PreAttacK}, \textsc{GANG-s} often misclassifies new users that receive many requests (Fig. \ref{fig:combevalcountry}).

\vspace{3mm}  
\section{Conclusion}

In this paper, we have studied a principled algorithmic approach to address what we call the early detection paradox: mainstream algorithms to detect fake accounts rely on the same behaviors they seek to prevent, such as fake accounts' friendships and the content they share with others. To overcome this paradox, we show some of the first distributional analyses of how fake (and real) accounts send and receive friend requests after joining a major social network, before they have made friends or shared content. We show that these friend request behaviors evoke a natural multi-class extension to the preferential attachment model of social network growth. We leverage this model to derive a new algorithm \textsc{PreAttacK}, and we show that in relevant problem instances, \textsc{PreAttacK} near-optimally approximates the posterior probability that a new user is fake. This approach also provides some of the first theoretic guarantees for fake account detection that do not rely on homophily assumptions. We conduct a variety of evaluations on the global Facebook network, and we consistently find that \textsc{PreAttacK} obtains informative classifications of new accounts before the median fake account succeeds in making a single friendship (i.e. \emph{accepted} friend request) with a real user. We note that, while impressive, \textsc{PreAttacK}'s AUC does not match state-of-the-art feature-based classifiers such as \textsc{DEC}, which \emph{eventually} obtains AUC$>$$0.98$ on the set of all active accounts by leveraging $\sim$$20$,$000$ user-features that describe users' friendships and shared content \citep{xudeep}. Instead, \textsc{PreAttacK} complements such methods by obtaining informative and interpretable early classifications before fake accounts can populate a user-feature vector, share content, or interact with others. 

\bibliographystyle{ACM-Reference-Format}
\bibliography{REFERENCES}


\appendix

\onecolumn
\clearpage
\section*{Appendix}

\section{Deferred analysis for instance-specific bounds}
\label{boundproof}

\paragraph{\textbf{Lower bound $f^F$}} We seek a worst-case factor $f^F \le \mathbf{\hat{P}_u}/\mathbf{P^*_u}$ that bounds \textsc{PreAttacK}'s underestimation of the posterior probability that $u$ is fake, which is useful in case $u$ is fake (Section \ref{sec:analysis}). The main difficulty is that a new user who sent/received requests before $u$ may have had both positive and negative effects (i.e. via its different edges) on $u$'s posterior. To sidestep the problem of trying all combinations of new users' latent labels, we bound the worst-case by supposing each new edge before $u$'s edges contained a \emph{unique} new `phantom' user whose latent label was the worst-case for its respective edge. Thus, any new user's edge before $u$'s edges that contained the same preexisting user as the one in $u$'s edge gets a \emph{fake} phantom user; any other such edge gets a \emph{real} phantom user. We compute the exact posterior (eqn. \ref{posterior}) using these phantom users' labels (in place of the latent ones) to obtain the desired bound. The probabilities for each of $u$'s observed edges become:
\begin{align}
P_{\nu_i+}^F &\coloneqq P[\nu_i|\ell_u \! = \! F,\cdot] = 
\frac{\alpha + \sum_{e_{x\rightarrow y} \in E_{i-1}}  \mathbbm{1}[y \! = \! \nu \wedge \ell_x \! = \! F] }{
\alpha|V| + \sum_{e_{x\rightarrow y} \in E_{i-1}}  \mathbbm{1}[ \ell_x \! = \! F]    }
\le
P_{\nu_i+,WCF}^F\coloneqq
\frac{\alpha + \sum_{e_{x\rightarrow y} \in \mathbf{E_0}}  \mathbbm{1}[y \! = \! \nu \wedge \ell_x \! = \! F] + \sum_{e_{x\rightarrow y} \in E_{i-1}\backslash E_0}  \mathbbm{1}[y \! = \! \nu]} {
\ \ \ \alpha|V| + \sum_{e_{x\rightarrow y} \in \mathbf{E_0}}  \mathbbm{1}[ \ell_x=F]  \ \ + \ \ \sum_{e_{x\rightarrow y} \in E_{i-1}\backslash E_0}  \mathbbm{1}[y=\nu] }
\\ %
P_{\nu_i+}^R &\coloneqq P[\nu_i|\ell_u \! = \! R,\cdot] = 
\frac{\alpha + \sum_{e_{x\rightarrow y} \in E_{i-1}}  \mathbbm{1}[y \! = \! \nu \wedge \ell_x \! = \! R] }{
\alpha|V| +  \sum_{e_{x\rightarrow y} \in E_{i-1}}  \mathbbm{1}[ \ell_x \! = \! R]  }
\ge 
P_{\nu_i+,WCF}^R \coloneqq
\frac{\alpha + \sum_{e_{x\rightarrow y} \in \mathbf{E_0}}  \mathbbm{1}[y \! = \! \nu \wedge \ell_x \! = \! R] }{
\alpha|V| +  \sum_{e_{x\rightarrow y} \in \mathbf{E_0}}  \mathbbm{1}[ \ell_x \! = \! R]  +  \sum_{e_{x\rightarrow y} \in E_{i-1}\backslash E_0}  \mathbbm{1}[y \! \not{=} \! \nu]}
\\ 
P_{\nu_i-}^F &\coloneqq P[\nu_i|\ell_u \! = \! F,\cdot] = \frac{ \alpha + \sum_{e_{x\rightarrow y} \in E_{i-1}} \mathbbm{1}[x \! = \! \nu \wedge \ell_y \! = \! F] }{
\alpha|V| +  \sum_{e_{x\rightarrow y} \in E_{i-1}}  \mathbbm{1}[ \ell_y \! = \! F]    }
\le
P_{\nu_i-,WCF}^F \coloneqq
 \frac{ \alpha + \sum_{e_{x\rightarrow y} \in \mathbf{E_0}} \mathbbm{1}[x \! = \! \nu \wedge \ell_y \! = \! F]  + \sum_{e_{x\rightarrow y} \in E_{i-1}\backslash E_0}  \mathbbm{1}[x=\nu] }{
\alpha|V| +  \sum_{e_{x\rightarrow y} \in \mathbf{E_0}}  \mathbbm{1}[ \ell_y \! = \! F]   
\ \ + \ \ \sum_{e_{x\rightarrow y} \in E_{i-1}\backslash E_0}  \mathbbm{1}[x=\nu] }
\\ %
P_{\nu_i-}^R &\coloneqq P[\nu_i|\ell_u \! = \! R,\cdot] =
\frac{\alpha + \sum_{e_{x\rightarrow y} \in E_{i-1}}  \mathbbm{1}[x \! = \! \nu \wedge \ell_y \! = \! R] }{
\alpha|V| + \sum_{e_{x\rightarrow y} \in E_{i-1}} \mathbbm{1}[ \ell_y \! = \! R]  }
\ge
P_{\nu_i-,WCF}^R\coloneqq
\frac{\alpha + \sum_{e_{x\rightarrow y} \in \mathbf{E_0}}  \mathbbm{1}[x \! = \! \nu \wedge \ell_y \! = \! R] }{
\alpha|V| +  \sum_{e_{x\rightarrow y} \in \mathbf{E_0}}  \mathbbm{1}[ \ell_y \! = \! R]  + \sum_{e_{x\rightarrow y} \in E_{i-1}\backslash E_0}  \mathbbm{1}[x\not{=}\nu] }
\end{align}


\begin{align}
\mathbf{\hat{P}_u}/\mathbf{P^*_u} &\coloneqq
\mathbf{\hat{P}_u} / \Big( \frac{
P_{\mathcal{N}_u^+}^F \cdot P_{\mathcal{N}_u^-}^F \cdot \pi
}{
P_{\mathcal{N}_u^+}^F \cdot P_{\mathcal{N}_u^-}^F \cdot \pi \ + \ P_{\mathcal{N}_u^+}^R \cdot P_{\mathcal{N}_u^-}^R \cdot (1-\pi)
}\Big )\\
&= 
\mathbf{\hat{P}_u} / \Big(
\frac{
\prod_{\nu_i \in \mathcal{N}_u^+}  P_{\nu_i+}^F \ \cdot \  \prod_{\nu_i\in \mathcal{N}_u^-}  P_{\nu_i-}^F \ \cdot \ \pi
}{
\prod_{\nu_i \in \mathcal{N}_u^+}  P_{\nu_i+}^F \ \cdot \ \prod_{\nu_i\in \mathcal{N}_u^-}  P_{\nu_i-}^F \ \cdot \ \pi \ \ + \ \ \prod_{\nu_i \in \mathcal{N}_u^+}  P_{\nu_i+}^R \ \cdot \  \prod_{\nu_i\in \mathcal{N}_u^-}  P_{\nu_i-}^R \ \cdot \  (1-\pi)
} \Big)
\\
&\ge f^F\coloneqq 
\mathbf{\hat{P}_u} / \Big(
\frac{
\prod_{\nu_i \in \mathcal{N}_u^+}  P_{\nu_i+,WCF}^F \ \cdot \  \prod_{\nu_i\in \mathcal{N}_u^-}  P_{\nu_i-,WCF}^F \ \cdot \ \pi
}{
\prod_{\nu_i \in \mathcal{N}_u^+}  P_{\nu_i+,WCF}^F \ \cdot \ \prod_{\nu_i\in \mathcal{N}_u^-}  P_{\nu_i-,WCF}^F \ \cdot \ \pi \ \ + \ \ \prod_{\nu_i \in \mathcal{N}_u^+}  P_{\nu_i+,WCF}^R \ \cdot \  \prod_{\nu_i\in \mathcal{N}_u^-}  P_{\nu_i-,WCF}^R \ \cdot \  (1-\pi)
}
\Big)
\label{posterior_boundF}
\end{align}
This expression for $f^F$ requires no knowledge of new users' latent labels, and conveniently, it can be computed during the same single pass through new edges that we use to compute \textsc{PreAttacK} with no penalty in asymptotic complexity (Note also that the expression for $f^F$ can be further factored as in eqn. \ref{posterior}).

\paragraph{\textbf{Upper bound $f^R$}} We also seek a worst-case overestimate factor $f^R \ge \mathbf{\hat{P}_u}/\mathbf{P^*_u}$ that bounds \textsc{PreAttacK}'s overestimation of the posterior probability that $u$ is fake, which is useful in case $u$ is real. Similar to before, we bound the worst-case by computing the exact posterior (eqn. \ref{posterior}) supposing each new edge before $u$'s edges contained a unique new `phantom' user whose latent label was the worst-case for $u$'s posterior:

\begin{align}
P_{\nu_i+}^F 
&\ge
P_{\nu_i+,WCR}^F\coloneqq
\frac{\alpha + \sum_{e_{x\rightarrow y} \in \mathbf{E_0}}  \mathbbm{1}[y=\nu \wedge \ell_x=F]  }{
\ \ \ \alpha|V| +  \sum_{e_{x\rightarrow y} \in \mathbf{E_0}} \big( \alpha+ \mathbbm{1}[ \ell_x=F] \big) \ \ + \ \ \sum_{e_{x\rightarrow y} \in E_{i-1}\backslash E_0}  \mathbbm{1}[y\not{=}\nu] }
\\%
P_{\nu_i+}^R 
&\le 
P_{\nu_i+,WCR}^R \coloneqq
\frac{\alpha + \sum_{e_{x\rightarrow y} \in \mathbf{E_0}}  \mathbbm{1}[y=\nu \wedge \ell_x=R] + \sum_{e_{x\rightarrow y} \in E_{i-1}\backslash E_0} \mathbbm{1}[y=\nu]}{
\alpha|V| + \sum_{e_{x\rightarrow y} \in \mathbf{E_0}} \mathbbm{1}[ \ell_x=R]  +  \sum_{e_{x\rightarrow y} \in E_{i-1}\backslash E_0}  \mathbbm{1}[y=\nu]}
\\
P_{\nu_i-}^F 
&\ge
P_{\nu_i-,WCR}^F \coloneqq
 \frac{ \alpha + \sum_{e_{x\rightarrow y} \in \mathbf{E_0}} \mathbbm{1}[x=\nu \wedge \ell_y=F] }{
\alpha|V| + \sum_{e_{x\rightarrow y} \in \mathbf{E_0}} \mathbbm{1}[ \ell_y=F]    
\ \ + \ \ \sum_{e_{x\rightarrow y} \in E_{i-1}\backslash E_0}  \mathbbm{1}[x\not{=}\nu]} 
\\%
P_{\nu_i-}^R 
&\le
P_{\nu_i-,WCR}^R\coloneqq
\frac{\alpha + \sum_{e_{x\rightarrow y} \in \mathbf{E_0}}  \mathbbm{1}[x=\nu \wedge \ell_y=R] + \sum_{e_{x\rightarrow y} \in E_{i-1}\backslash E_0}  \mathbbm{1}[x=\nu]}  {
\alpha|V| + \sum_{e_{x\rightarrow y} \in \mathbf{E_0}}  \mathbbm{1}[ \ell_y=R]  + \sum_{e_{x\rightarrow y} \in E_{i-1}\backslash E_0}  \mathbbm{1}[x=\nu] }
\end{align}

\begin{align}
\mathbf{\hat{P}_u}/\mathbf{P^*_u} &\coloneqq
\mathbf{\hat{P}_u} / \Big(
\frac{
\prod_{\nu_i \in \mathcal{N}_u^+}  P_{\nu_i+}^F \ \cdot \  \prod_{\nu_i\in \mathcal{N}_u^-}  P_{\nu_i-}^F \ \cdot \ \pi
}{
\prod_{\nu_i \in \mathcal{N}_u^+}  P_{\nu_i+}^F \ \cdot \ \prod_{\nu_i\in \mathcal{N}_u^-}  P_{\nu_i-}^F \ \cdot \ \pi \ \ + \ \ \prod_{\nu_i \in \mathcal{N}_u^+}  P_{\nu_i+}^R \ \cdot \  \prod_{\nu_i\in \mathcal{N}_u^-}  P_{\nu_i-}^R \ \cdot \  (1-\pi)
} \Big)
\\
&\le f^R\coloneqq
\mathbf{\hat{P}_u} / \Big(
\frac{
\prod_{\nu_i \in \mathcal{N}_u^+}  P_{\nu_i+,WCR}^F \ \cdot \  \prod_{\nu_i\in \mathcal{N}_u^-}  P_{\nu_i-,WCR}^F \ \cdot \ \pi
}{
\prod_{\nu_i \in \mathcal{N}_u^+}  P_{\nu_i+,WCR}^F \ \cdot \ \prod_{\nu_i\in \mathcal{N}_u^-}  P_{\nu_i-,WCR}^F \ \cdot \ \pi \ \ + \ \ \prod_{\nu_i \in \mathcal{N}_u^+}  P_{\nu_i+,WCR}^R \ \cdot \  \prod_{\nu_i\in \mathcal{N}_u^-}  P_{\nu_i-,WCR}^R \ \cdot \  (1-\pi)
}
\Big)
\label{posterior_boundR}
\end{align}
Note that $f^F$ is strictly decreasing, and $f^R$ strictly increasing in the number of new users' edges before $u$'s. 



\twocolumn

\section{Multi-Class \textsc{PreAttacK}}
\label{appendix:multiclass}
\textsc{PreAttacK} also applies to the case where we want to classify $k$$>$$2$ classes $\kappa\in\{1,\dots,k\}$ of fake users, such as sockpuppets, false news bots \citep{vosoughi2018spread}, etc. and each has different preferences in terms of existing users they seek to befriend. Computing \textsc{PreAttacK} in this case, we first compute $k$ conditional probabilities $\hat{P}_{\nu_i+}^\kappa$ of each observed edge that new user $u$ sends---one for each class $\kappa$---and also $k$ conditional probabilities $ \hat{P}_{\nu_i-}^\kappa$ of each observed edge that new user $u$ receives:

\begin{align}
\hat{P}_{\nu_i+}^\kappa &= \frac{\alpha + \sum_{e_{x\rightarrow y} \in \mathbf{E_0}}  \mathbbm{1}[y=\nu \wedge \ell_x=\kappa] }{
\alpha|V| + \sum_{e_{x\rightarrow y} \in \mathbf{E_0}}  \mathbbm{1}[ \ell_x=\kappa]    }\ ;\\ \ \ \ \ \ \hat{P}_{\nu_i-}^\kappa &= \frac{\alpha + \sum_{e_{x\rightarrow y} \in \mathbf{E_0}} \mathbbm{1}[x=\nu \wedge \ell_y=\kappa] }{
\alpha|V| + \sum_{e_{x\rightarrow y} \in \mathbf{E_0}}  \mathbbm{1}[ \ell_y=\kappa]    }
\end{align}

We use these (approximated) conditional probabilities to compute $k$ joint conditional probabilities $P_{\mathcal{N}_u^+}^\kappa$ of all $u$'s outgoing requests, and $k$ probabilities $P_{\mathcal{N}_u^-}^\kappa$ of all $u$'s incoming requests. Finally we compute $k$$-$$1$ (approximated) posterior probabilities---one for each class that could describe the new account (the $k$'th is implied). Here, the posterior probability that $u$ is a member of class $\kappa$ that has prior probability $\pi^\kappa$ is:

\begin{align}
\mathbf{\hat{P}^\kappa_u} 
&=
\frac{
\hat{P}_{\mathcal{N}_u^+}^\kappa \cdot \hat{P}_{\mathcal{N}_u^-}^\kappa \cdot \pi^\kappa
}{
\hat{P}_{\mathcal{N}_u^+}^\kappa \cdot \hat{P}_{\mathcal{N}_u^-}^\kappa \cdot \pi^\kappa \ + \ \displaystyle\sum_{\gamma=\{1,\dots,\kappa\}\backslash\kappa} ^ k \hat{P}_{\mathcal{N}_u^+}^\gamma \cdot \hat{P}_{\mathcal{N}_u^-}^\gamma \cdot \pi^\gamma
}
\ \ \\&= \ \ 
\Big( 1\ + \ (P_{\mathcal{N}_u^+}^\kappa \cdot P_{\mathcal{N}_u^-}^\kappa \cdot \pi^k)^{-1} \displaystyle\sum_{\gamma=\{1,\dots,\kappa\}\backslash\kappa} ^ k \hat{P}_{\mathcal{N}_u^+}^\gamma \cdot \hat{P}_{\mathcal{N}_u^-}^\gamma \cdot \pi^\gamma \Big)^{-1}
\label{posteriormulticlass}
\end{align}

\section{Adversarial robustness in practice: additional discussion}
\label{appendix:selectionbias}
We also note a relationship between our work and the very recent discussions in the fake account and fake news detection community that have highlighted causal considerations, and in particular, unobserved confounding and related biases (see e.g. \citet{cheng2021causal}).  Specifically, algorithmic approaches to fake accounts (or fake news) will suffer from bias to the extent that they leverage users' counts of fake friends (or fake news articles shared) without accounting for the fact that some users had more exposure to fake accounts (e.g. by receiving more friend requests from fakes) than others \emph{a priori}, and some users had more exposure to real accounts than others \emph{a priori}. Such bias can explicitly accrue when, for example, algorithms increase the posterior belief that a new account is real by the same amount for each real account she befriends, failing to account for the fact that different users have varying propensities to receive friend requests from fake accounts.
It is also well-known that adversaries can leverage this bias to avoid detection. For example, an adversary might avoid detection by mainstream algorithms by sending thousands of friend requests to real users in the hope of befriending a `normal' number of real friends, knowing that certain algorithms are blind to rejected friend requests and unable to account for the fake account's unduly high \emph{a priori} exposure to real users.     Our distributional analyses in Section \ref{intro} confirm this `varying exposure' phenomenon for our problem instance. 
 Sophisticated adversaries may also attempt to learn the subset of real users who accept friend requests indiscriminately, or they may strategically delete certain friendships after making them in order to manipulate detection algorithms (see e.g. \citep{xue2013votetrust, torres2022manipulating, xudeep}).

    Whereas several new research streams seek to address these sources of bias via propensity scoring and other inference techniques, \textsc{PreAttacK} sidesteps them entirely by aggregating over all friend requests (rather than the subset that are accepted). More importantly, \textsc{PreAttacK} does not require inferential propensity scoring corrections to address bias from to the fact that different users have varying exposure to fake accounts, as we are able to use social network data to explicitly compute each existing user's `propensity' to receive a friend request from (or send a request to) fake and real accounts, which we compute as each user's preferential attachment probabilities, $\hat{P}_{\nu+}^F, \hat{P}_{\nu+}^R, \hat{P}_{\nu-}^F, \hat{P}_{\nu-}^R.$ In this way, \textsc{PreAttacK} may be seen as robust to prevalent confounding bias vulnerabilities that are common among mainstream approaches in this application domain.

\bigskip
\section{Details of Homophily Benchmark}
\label{appendix:evaluations}

\textbf{Homophily} is equivalent to \textsc{PreAttacK++} with $E_0=\varnothing$. Thus, all  request probabilities in \textsc{Homophily} can be summarized by an $8$-tuple, including $4$ probabilities $\alpha_{\ell_u\rightarrow\ell_\nu}^+$ that a new $\{fake,real\}$ \emph{sends} a request to a preexisting $\{fake,real\}$ and $4$ probabilities $\alpha_{\ell_\nu\rightarrow\ell_u}^-$ that a new $\{fake,real\}$ \emph{receives} a request from a preexisting $\{fake,real\}$. Here, \hyperref[kCDPA]{$k$CDPA} reduces to a $2$-class directed Stochastic Block Model (SBM) \citep{wang1987stochastic}.

\medskip
\section{Details of benchmark algorithms}
\label{appendix:benchmarks}
Section \ref{sec:eval} compares \textsc{PreAttacK} and its variants to four state-of-the-art benchmarks. For each, we use their paper's code and parameters:

\vspace{-2mm}
\paragraph{\textbf{GANG}} \textsc{GANG}  \cite{wang2017gang} is a recent algorithm that leverages directed edges (requests) in a belief-propagation framework. We consider two variants: \textsc{GANG-s}, which uses the directed network of friendship requests \emph{sent}, and \textsc{GANG-r}, which uses the directed network of requests \emph{received}. This allows us to test \textsc{GANG}'s performance when beliefs about the authenticity of a new user flow from senders to receivers, or alternatively, from receivers to senders. As with \textsc{SybilBelief}, we set parameters $\{\theta^+, \theta^-, \theta\}$ to $\{0.9, 0.1, 0.5\}$ per \cite{wang2017gang}.

\vspace{-2mm}
\paragraph{\textbf{SybilRank}} \textsc{SybilRank} \cite{yang2012sybilrank} is currently the most widely used random walk based algorithm. \textsc{SybilRank} runs on the network of accepted friend requests and set of known real users. As in \cite{yang2012sybilrank}, we run \textsc{SybilRank} for log2$(|V|)$ iterations.

\vspace{-2mm}
\paragraph{\textbf{SybilBelief}}\textsc{SybilBelief}  \cite{gong2014sybilbelief} is a loopy belief propagation algorithm that is widely used in state-of-the-art applications. \textsc{SybilBelief} uses the network of accepted friend requests and both known real users and fakes. As in \cite{gong2014sybilbelief}, we run \textsc{SybilBelief} with edge weights of $0.9$ and set $\{\theta^+, \theta^-, \theta\}$ to $\{0.9, 0.1, 0.5\}$.

\vspace{-2mm}
\paragraph{\textbf{SybilSCAR}}\textsc{SybilSCAR}  \cite{wang2018structure} is a recent algorithm that uses the graph of accepted requests and both known real users and fakes. We run both versions: \textsc{SybilSCAR-C} with weights equal to half the inverse of the avg. degree per \cite{wang2018structure}, and user-degree weighted \textsc{SybilSCAR-D}. Each point in Figs. \ref{fig:RMevalcountry} \& \ref{fig:combevalcountry} reports the higher of their two AUC's. Per \citep{wang2018structure}, we set $\{\theta^+, \theta^-, \theta\}$ to $\{0.6, 0.4, 0.5\}$, and $\delta$$ = $$10^{-3}$.


\end{document}